\documentclass[12pt,a4paper]{article}
\usepackage{amssymb,amsmath,amsthm}
\usepackage{graphicx}
\usepackage{epsfig}
\usepackage{geometry}
\usepackage{tikz}
\usetikzlibrary{shapes}
\usepackage{setspace}
\usepackage{tikz}
\usetikzlibrary{arrows}
\usepackage{float}

\tikzset{every picture/.style={line width=0.5mm}}

\newcommand\stringdiagram[1]{
	\begin{center}
		\begin{tikzpicture}[	scale=0.75,
				  		>=latex,
						triangle/.style={regular polygon, regular polygon sides=3},
						inverted triangle/.style={triangle, shape border rotate=180}
						]
			#1
		\end{tikzpicture}
	\end{center}
}

\newcommand{\G}{\mathcal G}
\renewcommand{\H}{\mathcal H}

\renewcommand{\P}{\mathbf P}
\newcommand{\C}{\mathbf C}
\newcommand{\E}{\mathbf E}
\newcommand{\R}{\mathbb R}
\newcommand{\id}{\operatorname{id}}

\newcommand\tensor\otimes

\title{Compositionality and String Diagrams\\for Game Theory}
\author{Jules Hedges\\
\scriptsize School of Electronic Engineering and Computer Science, Queen Mary University London\\
Evguenia Shprits\\
\scriptsize Department of Economics, University of Mannheim\\
Viktor Winschel\\
\scriptsize Department of Management, Technology and Economics, ETH Z\"urich\\
Philipp Zahn\\
\scriptsize Department of Economics, University of St. Gallen
}
\date{\today}
\newtheorem{definition}{Definition}
\begin{document}
\maketitle

\begin{abstract}
We introduce string diagrams as a formal mathematical, graphical language to represent, compose, program and reason about games. 
The language is well established in quantum physics, quantum computing and quantum linguistic with the semantics given by category theory.
We apply this language to the game theoretical setting and show examples how to use it for some economic games
where we highlight the compositional nature of our higher-order game theory.
\end{abstract}
\newpage
\setcounter{tocdepth}{2}
\tableofcontents
\section{Introduction}  

%
%
%

Within a few years, a junior software engineer can grow from studying small, textbook example of programs, to working in a team with 100 other software engineers on a product containing tens of millions of lines of code.
Why do other disciplines not achieve this level of scalability?
The most fundamental answer is \emph{compositionality}.

Compositionality is the principle that a system should be built by joining together (or `composing') simpler systems, and that all reasoning about the system should be done recursively on the way that it was built.
Just as importantly, it should be possible for a designer to use a sub-system having only an abstract view of it, rather than knowing every detail about it.
For example, a function in a programming language can be used knowing only its specification (for example, `sort a list'), but not its implementation.
This means, for example, that the implementation can be changed (such as changing from merge sort to quicksort), without any other code needing to be changed.

The same holds for a physical component used by an engineer: a component, for example a fluid pump, could be changed for a different component with the same characteristics such as power and flow rate.
The same again holds for a human organisation such as a company: for example, an employee can assume that the electricity bill is being paid without knowing the process, and they need not even be aware if the electricity provider is changed.

Most disciplines, of course, are not compositional, and hence are not able to be applied on a large scale.
To pick one example, game theory is not compositional: there is generally no meaningful sense in which a game can be built by joining smaller games together.
Recently it has been shown that it is possible to build a foundation for game theory that is inherently compositional, however.
At small scales this appears to be unnecessary complexity, but it will distinguish itself when working at large scales, the economic equivalent of tens of millions of lines of code.

The purpose of this paper is to introduce the theory, making no strong assumptions about mathematical background, and to give economically-motivated examples.
The heaviest parts of the theory can be entirely avoided by using so-called \emph{string diagrams}.

\begin{figure}[ht!]
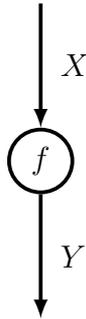

\stringdiagram{
	\node (X) at (0,6) {};
	\node [circle, draw] (f) at (0,3) {$f$};
	\node (Y) at (0,0) {};
	\draw [->] (X) to [out = -90, in = 90] node [right = 0.1cm, midway] {$X$} (f);
	\draw [->] (f) to [out = -90, in = 90] node [right = 0.1cm, midway] {$Y$} (Y);
}
\caption{Basic String Diagram}
\label{fig:bas}
\end{figure}

String diagrams are a simple but powerful graphical algebra that has been recently developed by a multi-disciplinary community including researchers from theoretical physics, computer science, linguistics and pure mathematics.
The basic components are \emph{strings} that represent state spaces, and \emph{nodes} 
that represent processes or transformations. 
A process $f$ that transforms a state space $X$ into a state space $Y$ is represented in Figure \ref{fig:bas}.

Notice that  in a string diagram time conventionally flows downwards. 
The diagram (or theory, or syntax) in Figure \ref{fig:bas} has many possible interpretations (or models, or semantics). 
For example, we could interpret it in the following ways, among others:
\begin{itemize}
\item The simplest interpretation is that $X$ and $Y$ are sets and $f$ is a function $f: X \to Y$.
\item Alternatively we can interpret $X$ and $Y$ as sets but $f \subseteq X \times Y$ as a relation.
\item In quantum mechanics $X$ and $Y$ are Hilbert spaces and $f$ is a linear operator or matrix.
\item If we are interested in random processes, we can interpret $X$ and $Y$ as measurable spaces and $f$ as a measurable function.
\item In computer science we can interpret $X$ and $Y$ as datatypes and $f$ as a computer program with input type $X$ and output type $Y$.
\item In logic, we can interpret $X$ and $Y$ as logical formulas and $f$ as a proof that the conclusion $Y$ follows from the hypotheses $X$.
\end{itemize}
The power of string diagrams arises from two facts: the graphical language is simple and intuitive and yet the semantics underlying it can be very complex, so that the difficulty is hidden behind the scenes. For example, string diagrams are being heavily used to study quantum-cryptographic protocols, in which reasoning with Hilbert spaces, complex matrices and Dirac notation is reduced to trivial operations on string diagrams. There is no trick because the hard work instead goes into coherence
theorems that guarantee that the meaning of a string diagram is invariant under their topological deformations. However, once these theorems have been proven, the string language can be used to prove theorems without the need to keep track of the mathematical details. 

This paper provides an introduction to string diagrams as a representation of strategic games developed in \cite{Ghani_Hedges2016}. We illustrate how string diagrams can be used to model games and - in contrast to the graphical representation of the game tree - to generate new games with two basic operations: parallel and sequential composition. There is also a way to modularize games in string diagrams by boxing them and thus compose hierarchical games which can be reused in different places. 

This paper is meant as an introduction for economists without the technical background in theoretical computer science. We present only the machinery that is absolutely necessary and mostly focus on simple, pedagogical examples. The basis for the current paper is laid out in \cite{Hedges_et_al_2015_decisions,Hedges_et_al_2015_games} where we introduce a new representation of decision problems and games based on higher-order functions. 
Starting from this, \cite{Ghani_Hedges2016} shows that the framework can be extended to form a category and that this category fulfils certain axioms 
which in turn guarantee the existence of the diagrammatic language.\footnote{For the interested reader, the appendix contains condensed information on the underlying mathematical treatment. We refer to \cite{Ghani_Hedges2016} for the details.}  

The paper is organized as follows. In the next section \ref{sec:strings} we give a brief background on the origin of string diagrams and their applications in physics, linguistics and computer science.
In section \ref{sec:opengames} we introduce the mathematical structure that is necessary to translate games into string diagrams.
We define the notion of open games, the operators on them and the graphical algebraic operations.
In the rest of the paper in section \ref{sec:games}, we illustrate the use of string diagrams with several examples.
\section{String Diagrams}\label{sec:strings}
String diagrams \emph{are} literally algebraic expressions only with a less familiar notation and manipulating them is the same as doing the equivalent algebraic operations.
This is opposed to, for instance,  game trees but also influence diagrams \cite{Koller2003} that extend Bayesian network \cite{Pearl1988,Pearl2009}. 
These graphical formalisms are \emph{not} algebraic expressions but a graphical representations or metaphors for some underlying mathematics.
\subsection{String Diagrams in the Sciences}

String diagrams are a graphical calculus that can be used to visualise information flow in a type of algebraic structure called a monoidal category.
For an excellent introduction to string diagrams and how they connect several subjects, see \cite{Baez2009}.

Their earliest appearance may be Penrose's graphical tensor notation in \cite{penrose1971applications}, with another precursor being Girard's proof nets for linear logic \cite{girard87}, and the mathematical foundations were formalised in \cite{joyal91}.
They became well known through the work of Samson Abramsky, Bob Coecke and others on quantum information theory \cite{abramsky04}, and later through the work of Bob Coecke, Mehrnoosh Sadrzadeh and others on distributional semantics in linguistics \cite{coecke13}.
These are large enough areas to be covered by books, \cite{Coecke2011} and \cite{Heunen2013}.
String diagrams are also being applied in other areas such as bialgebra \cite{bonchi14}, computability \cite{pavlovic13} and probability theory \cite{fong12}.
Finally, the many variants of the string diagram language appearing in different application areas are surveyed in \cite{selinger11}.

As is the case in quantum physics and linguistics, string diagrams can be used in game theory to visualise \emph{information flow}.
This visualisation is a separate issue to compositionality, although being able to compose string diagrams is a crucial requirement.

The purpose of string diagrams varies by discipline. In quantum information theory, the emphasis is generally on the ability of string diagrams to reduce complex calculations to trivial topological deformation, a point made forcibly in \cite{Coecke2005}.
In linguistics, there is more emphasis on the use of string diagrams as a device for visualising the logical structure of sentences, whereas the underlying categorical structure is used in a more formal way by considering functorial semantics.

In game theory, we will similarly emphasise string diagrams as a visualisation tool.
However, more so than in linguistics, the algebraic expressions denoting even simple games can be quite complicated, and we will make use of string diagrams as a \emph{tool for making definitions}.
For the games towards the end of this paper, the procedure of converting the string diagram into its underlying meaning is quite involved.
However, this work is entirely mechanical and could be automated.

\subsection{Mathematical Backgrounds}
\begin{figure}[H]
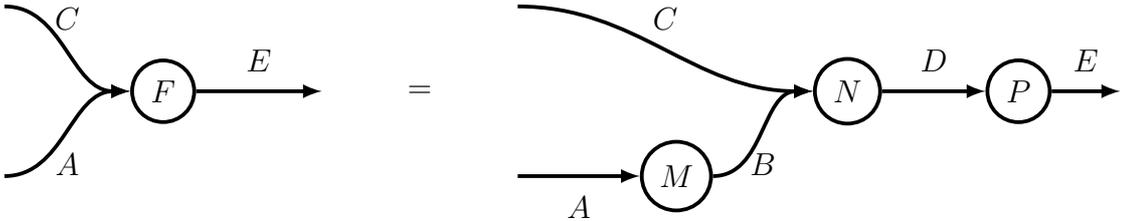

\stringdiagram{
	\node (C1) at (0,3){};
	\node (A1) at (0,0){};
	\node (F1) [circle, draw] at (3,1.5){$F$};
	\node (E1) at (6,1.5){};
	\draw [->] (C1) to [out = 0, in = 180] node [above = 0.1cm, midway] {$C$} (F1);
	\draw [->] (A1) to [out = 0, in = 180] node [below = 0.1cm, midway] {$A$} (F1);
	\draw [->] (F1) to [out = 0, in = 180] node [above = 0.1cm, midway] {$E$} (E1);

	\node (e1) at (7.5,1.5){$=$};

	\node (C2) at (9,3){};
	\node (A2) at (9,0){};
	\node (M2) [circle, draw] at (12,0){$M$};
	\node (N2) [circle, draw] at (15,1.5){$N$};
	\node (P2) [circle, draw] at (18,1.5){$P$};
	\node (E2) at (20,1.5){};
	\draw [->] (C2) to [out = 0, in = 180] node [above = 0.1cm, midway] {$C$} (N2);
	\draw [->] (A2) to [out = 0, in = 180] node [below = 0.1cm, midway] {$A$} (M2);
	\draw [->] (M2) to [out = 0, in = 180] node [below = 0.1cm, midway] {$B$} (N2);
	\draw [->] (N2) to [out = 0, in = 180] node [above = 0.1cm, midway] {$D$} (P2);
	\draw [->] (P2) to [out = 0, in = 180] node [above = 0.1cm, midway] {$E$} (E2);
}
\caption{Basic String Diagram Operations}
\label{fig:string_matrix}
\end{figure}

The language of string diagrams extends standard notation into a graphical one.
The graphical notation appeared  in \cite{penrose1971applications} for the first time as operator diagrams.
It was then used for tensor products of linear operators.
We can combine the linear maps
$M:A\rightarrow B, N:B\otimes C\rightarrow D, P:D\rightarrow E$ 
on finite dimensional vector spaces $A,B,C,D,E$
into the linear map $F:A\otimes C\rightarrow E$,
that is given by $F=P\circ N \circ (M\otimes id_{C})$ in a functional notation and by
$f_{m,ik}=\sum_{j}\sum_{l}p_{m,l}n_{l,jk}m_{j,i}$ as a summation over matrix indices of
$M=(m_{j,i}),N=(n_{l,jk}),P=(p_{m,l}),F=(f_{m,ik})$.
In Einstein notation we have
$F^{ik}_m=P^l_m N^{jk}_l M^i_j$
with column indices in superscripts and row indices as subscripts.
A string diagram representation is given in Figure \ref{fig:string_matrix}.

\begin{figure}[H]
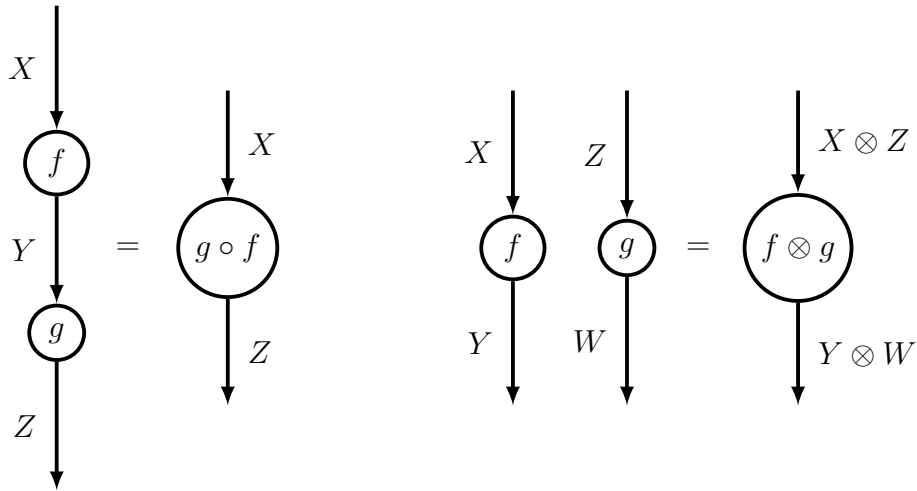

\stringdiagram{
	\node (X1) at (0,9) {};
	\node [circle, draw] (f1) at (0,6) {$f$};
	\node [circle, draw] (g1) at (0,3) {$g$};
	\node (Z1) at (0,0) {};
	\draw [->] (X1) to [out = -90, in = 90] node [left = 0.1cm, midway] {$X$} (f1);
	\draw [->] (f1) to [out = -90, in = 90] node [left = 0.1cm, midway] {$Y$} (g1);
	\draw [->] (g1) to [out = -90, in = 90] node [left = 0.1cm, midway] {$Z$} (Z1);

	\node (e1) at (1.25,4.5) {$=$};

	\node (X2) at (3,7.5) {};
	\node [circle, draw] (gf2) at (3,4.5) {$g\circ f$};
	\node (Z2) at (3,1.5) {};
	\draw [->] (X2) to [out = -90, in = 90] node [right = 0.1cm, midway] {$X$} (gf2);
	\draw [->] (gf2) to [out = -90, in = 90] node [right = 0.1cm, midway] {$Z$} (Z2);

	\node (X3) at (8,7.5) {};
	\node [circle, draw] (f3) at (8,4.5) {$f$};
	\node (Y3) at (8,1.5) {};
	\draw [->] (X3) to [out = -90, in = 90] node [left = 0.1cm, midway] {$X$} (f3);
	\draw [->] (f3) to [out = -90, in = 90] node [left = 0.1cm, midway] {$Y$} (Y3);
	\node (Z4) at (10,7.5) {};
	\node [circle, draw] (g4) at (10,4.5) {$g$};
	\node (W4) at (10,1.5) {};
	\draw [->] (Z4) to [out = -90, in = 90] node [left = 0.1cm, midway] {$Z$} (g4);
	\draw [->] (g4) to [out = -90, in = 90] node [left = 0.1cm, midway] {$W$} (W4);
	
	\node (e4) at (11.25,4.5) {$=$};

	\node (XZ5) at (13,7.5) {};
	\node [circle, draw] (fg5) at (13,4.5) {$f\tensor g$};
	\node (YW5) at (13,1.5) {};
	\draw [->] (XZ5) to [out = -90, in = 90] node [right = 0.1cm, midway] {$X\tensor Z$} (fg5);
	\draw [->] (fg5) to [out = -90, in = 90] node [right = 0.1cm, midway] {$Y\tensor W$} (YW5);

}
\caption{Basic String Diagram Operations}
\label{fig:basstring}
\end{figure}

There are two operations that all of these areas have in common, which correspond to the two ways of joining a pair of string diagrams together: either end-to-end or side-by-side as can be seen in Figure \ref{fig:basstring}.

The purpose of a string diagram is to represent an element of a particular kind of algebraic structure called a \emph{category}, or more specifically a \emph{symmetric monoidal category}. We will call the thing denoted by a string diagram its meaning, although semantics or denotation would be more usual. The meaning of a string diagram is calculated systematically in terms of the meaning of sub-diagrams using the algebraic operations of a category, starting from a collection of atomic components whose meaning is given. In particular we can join two string diagrams in two basic ways: either side-by-side by spacial juxtaposition, or end-to-end by actually joining the wires together. These correspond to algebraic operations called tensor product and categorical composition. An important example of a category used in quantum mechanics is the algebra of complex-valued matrices, where tensor product is the usual tensor product of matrices, and categorical composition is matrix multiplication. Thus if we begin with graphical components representing certain known matrices, we can calculate the matrix denoted by a string diagram using tensor product and multiplication. In game theory these two operations correspond roughly to simultaneous composition and sequential composition of games.

The crucial mathematical facts that make graphical reasoning a powerful tool are the so-called \emph{coherence theorems}, which guarantee that the meaning of a string diagram is invariant under \emph{topological deformations}. Put simply, there is no need to take care exactly how a diagram is drawn, but the only thing that matters is its topology - the source, destination and direction of each arrow. In quantum theory some quite nontrivial reasoning can be reduced to trivial graphical manipulation of string diagrams, using the coherence theorems as a bridge between the worlds of physics and topology. 

\section{Open games}\label{sec:opengames}

In this section we define the units of the games and their string diagrammatic representation.
We call these units \emph{open games} and they constitute the building blocks and at the same time are the composed games.
Accordingly we then go on and define the operations to compose string diagrams and define their mathematical meanings.
This constitutes the \emph{semantics} of open games, which makes our use of string diagrams into a genuinely formal logical language, rather than merely a suggestive notation.

We will focus on graphical reasoning and intuition and include most of the usual formal algebraic formulations in the appendix.

\begin{figure}[H]
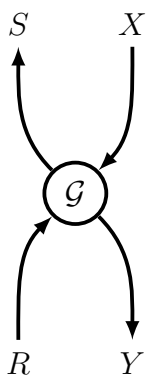

\stringdiagram{
	\node (X) at (2, 6) {$X$}; \node (Y) at (2, 0) {$Y$};
	\node (R) at (0, 0) {$R$}; \node (S) at (0, 6) {$S$};
	\node [circle, draw] (G) at (1, 3) {$\G$};
	\draw [->] (X) to [out=-90, in=45] (G); \draw [->] (G) to [out=-45, in=90] (Y);
	\draw [->] (R) to [out=90, in=-135] (G); \draw [->] (G) to [out=135, in=-90] (S);
}
\caption{Open Game}
\label{fig:string}
\end{figure}

A general open game is expressed by a string diagram in Figure \ref{fig:string}.
Equivalently, this is expressed by the algebraic notation
\[ \G : X \otimes S^* \to Y \otimes R^* \]
where the star attached to $S^*$ and $R^*$ hints that causality flows backwards in time indicated by the directions of the arrows.
Here $X$, $Y$, $R$ and $S$ are (in general different) sets, and serve different roles.
This notation is exactly analogous to multilinear algebra and quantum mechanics: if $X$, $Y$, $R$ and $S$ are finite-dimensional Hilbert spaces, $-^*$ is the duality of Hilbert spaces and $\otimes$ is the tensor product, then the string diagram representing a linear operator $\G : X \otimes S^* \to Y \otimes R^*$ is exactly Figure \ref{fig:string}.

The open game $\G$ may in general be an aggregate containing a mixture of computations or functions, players or games.
In degenerate cases, $\G$ may be only a single player or a single function, such as a utility function.
This is an important mathematical fact about open games: all components of game, including players, utility functions and aggregates thereof, are uniformly represented by the same kind of mathematical object, namely open games, and the same operations can be uniformly applied to them.
We can work with an open game knowing only its \emph{type}, which in the basic case is $\G : X \otimes S^* \to Y \otimes R^*$, without knowing or caring whether it represents a player, a utility function, or a complicated aggregate thereof.

The roles of the sets $X$, $Y$, $R$ and $S$ are as follows:
\begin{itemize}
	\item $X$ is the set of \emph{histories}, or \emph{observations} that can be made by $\G$. In general, players inside $\G$ may condition their strategy on $X$. However, if $\G$ is an aggregate, not all players in $\G$ may be perfectly informed about $X$.
	\item $Y$ is the set of \emph{choices} that $\G$ can make, which can be observed by later players. In case of $\G$ being a computation $Y$ is a function value or output.
	\item $R$ is the set of \emph{outcomes}, which is what the players inside $\G$ are acting to optimise. In most of the examples in this paper we will take $R$ to be the set of real numbers, in which case it represents \emph{utility} or \emph{payoff}. 
	However, our mathematical framework allows for more general possibilities, see \cite{Hedges_et_al_2015_games}.
	\item The set $S$ is the hardest part of the definition to understand: we call it \emph{coutility}. It represents utility `generated by' an open game, which is optimised by some player external to $\G$. 
	Players never generate coutilities, but outcome functions do. Coutilities are also used in repeated games in order to ``communicate'' utilities from later stages to early stages.
\end{itemize}

\subsection{The Semantics of Open Games}\label{sec:Semantics_Open}

The meaning of the string diagram in Figure \ref{fig:string} is a 4-tuple $\G = (\Sigma_\G, \P_\G, \C_\G, \E_\G)$, where:
\begin{itemize}
	\item $\Sigma_\G$ is the set of \emph{strategy profiles} of $\G$
	\item $\P_\G : \Sigma_\G \times X \to Y$ is the \emph{play function} of $\G$; it takes a strategy profile $\sigma \in \Sigma_\G$ and an observation $x \in X$, and it `plays' the strategy on the observation to produce a choice $\P_\G (\sigma, x) \in Y$
	\item $\C_\G : \Sigma_\G \times X \times R \to S$ is called the \emph{coplay function}; this can be viewed as a purely technical construction, and it will be used only in the appendix
	\item $\E_\G : X \times (Y \to R) \to \mathcal P (\Sigma_\G)$ is the \emph{equilibrium function}; it gives the subset of equilibrium strategies (technically, pure-strategy subgame perfect equilibria) of $\G$, given a \emph{context} consisting of an observation $x \in X$ and a function $k : X \to Y$.
\end{itemize}

This function $k : X \to Y$, in some ways, is the most important part of the whole definition and the secret to achieving compositionality.
It is an example of a \emph{continuation}, a concept in computer science made famous by \cite{Abelson1996}. 
In programming languages the purpose of a continuation is to represent the `future' of a computed value, in particular how the value will be used by the program's calling environment. In game theory the idea is similar: the continuation represents the way in which a choice will be used by an open game's `calling environment' to calculate an outcome.
Crucially, this function can encapsulate arbitrary complexity, including the choices made by other players with fixed strategies.

Every string diagram appearing in this paper denotes such a tuple $(\Sigma_\G, \P_\G, \C_\G, \E_\G)$, given this data for the basic components and the methods used to compose them, which are given in the appendix.
The actual calculation is often tedious and error-prone, but it is easy to automate using a high-level programming language such as Haskell, and in practice we already use a prototype implementation to check our calculations.

\subsection{String Repetition and Omission}

The four strings which enter and leave a string diagram, labelled by sets such as $X$, $Y$, $R$, $S$, will be called \emph{open ports}.
They are the channels through which an open game may communicate with its external environment in various ways.
This is the motivation of the term `open game', meaning `open to its environment'.

\begin{figure}[H]
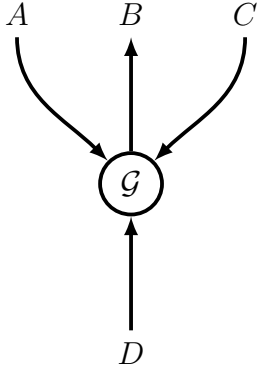

\stringdiagram{
	\node (A) at (0, 6) {$A$}; \node (B) at (2, 6) {$B$}; \node (C) at (4, 6) {$C$};
	\node [circle, draw] (G) at (2, 3) {$\G$}; \node (D) at (2, 0) {$D$};
	\draw [->] (A) to [out=-90, in=135] (G); \draw [->] (G) to (B); \draw [->] (C) to [out=-90, in=45] (G);
	\draw [->] (D) to (G);
}
\caption{Variant string diagram}
\label{fig:string-variant}
\end{figure}

In general, a string diagram may have some of these strings omitted or repeated.
For example, the string diagram in Figure \ref{fig:string-variant}
corresponds to the algebraic expression
\[ \G : A \otimes B^* \otimes C \to D^* \]
(and again, this is the same as the corresponding expression and string diagram in multilinear algebra).

In order to give the 4-tuple corresponding to this diagram, we need to reduce it to the previous case with exactly 4 strings.
We do this as follows:
\begin{itemize}
	\item Each of the 4 sets in the previous section is set to the cartesian product of all of the sets entering or leaving the string diagram in the same direction
	\item If there are no strings entering or leaving in some direction, we use the dummy set $I = \{ \bullet \}$ containing one canonical element.
\end{itemize}

In the example of Figure \ref{fig:string-variant} this is $X = A \times C$, $Y = I$, $R = D$ and $S = B$; thus in the tuple $(\Sigma_\G, \P_\G, \C_\G, \E_\G)$ we have
\begin{itemize}
	\item $\P_\G : \Sigma_\G \times A \times C \to I$ (and there is only one such function, namely the one that always returns $\bullet$)
	\item $\C_\G : \Sigma_\G \times A \times C \times D \to B$
	\item $\E_\G : A \times C \times D \to \mathcal P (\Sigma_\G)$ (where we have used the fact that functions $I \to D$ are in bijection with elements of $D$)
\end{itemize}

\begin{figure}[H]
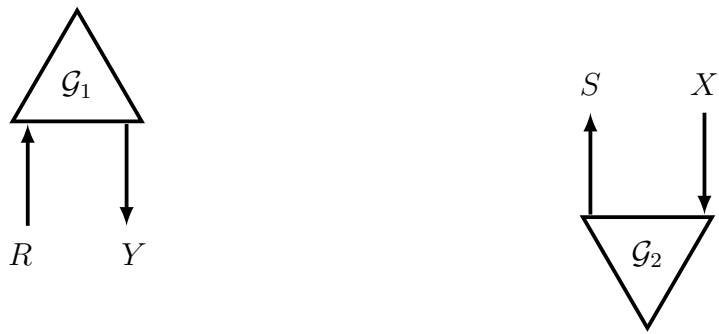

\stringdiagram{
	\node [triangle, draw] (G1) at (1, 3) {$\G_1$};
	\node (Y) at (2, 0) {$Y$}; \node (R) at (0, 0) {$R$};
	\draw [->] [transform canvas = {xshift= 0.1cm}] (0,0.5) to (0,2.3);
	\draw [->] [transform canvas = {xshift= 1.4cm}] (0,2.3) to (0,0.5);

	\node (X) at (12, 3) {$X$}; \node (S) at (10, 3) {$S$};
	\node [inverted triangle, draw] (G2) at (11, 0) {$\G_2$};
	\draw [->] [transform canvas = {xshift= 7.5cm}] (0,0.7) to (0,2.5);
	\draw [->] [transform canvas = {xshift= 9cm}] (0,2.5) to (0,0.7);
}
\caption{One-sided string diagrams}
\label{fig:triangles}
\end{figure}

If a string diagram has no strings at all (either entering or leaving) on one side, then this is denoted algebraically by $I$.
It is traditional in the string diagrams literature that such a string diagram is denoted by a triangle rather than a circle (but this is nothing more than a graphical convention, and is not reflected in the mathematics).
For example, the two string diagrams in Figure \ref{fig:triangles}
denote open games of type $\G_1 : I \to Y \otimes R^*$ and $\G_2 : X \otimes S^* \to I$.

\subsection{Open Game Units}\label{sec:Open_game_units}

We will now introduce the basic components from which we build our string diagrams.
The tuples $(\Sigma, \P, \C, \E)$ corresponding to each of these is given in the appendix.

\subsubsection{Players}

\begin{figure}[H]
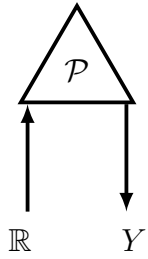

\stringdiagram{
	\node [triangle, draw] (P) at (1, 3) {$\mathcal P$};
	\node (Y) at (2, 0) {$Y$}; \node (R) at (0, 0) {$\R$};
	\draw [->] [transform canvas = {xshift= 0.1cm}] (0,0.5) to (0,2.4);
	\draw [->] [transform canvas = {xshift= 1.4cm}] (0,2.4) to (0,0.5);

}
\caption{Utility-maximising player making no observation}
\label{fig:another-player}
\end{figure}

The first family of components we consider are \emph{players}.
A player $\mathcal P$ who makes a choice in $Y$ in order to optimise a value in $R$, after observing a value in $X$, is shown in Figure \ref{fig:player}.
\begin{figure}[H]
\stringdiagram{
	\node (X) at (1, 6) {$X$}; \node (Y) at (2, 0) {$Y$}; \node (R) at (0, 0) {$R$};
	\node [circle, draw] (P) at (1, 3) {$\mathcal P$};
	\draw [->] (X) to (P); \draw [->] (P) to [out=-45, in=90] (Y); \draw [->] (R) to [out=90, in=-135] (P);
}
\caption{Player}
\label{fig:player}
\end{figure}
The corresponding algebraic type is
\[ \mathcal P : X \to Y \otimes R^* \]
In maximum generality a player is defined by a multi-valued selection function \cite{Hedges_et_al_2015_games}, but we can restrict to the classical case when $R$ carries a rational preference relation, or $R = \R$ is utility which the player acts to maximise.
If the player makes no observation then $X = I$, and the string diagram is shown in Figure \ref{fig:another-player} with the type being $\mathcal P : I \to Y \otimes \R^*$.

\subsubsection{Computations}

\begin{figure}[H]
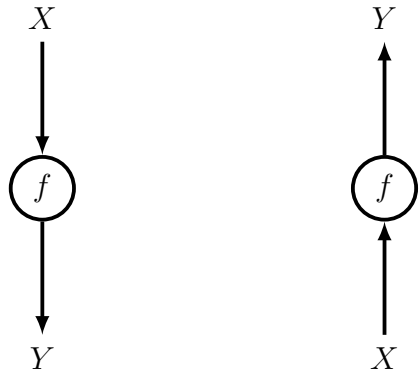

\stringdiagram{
	\node (X1) at (0, 6) {$X$}; \node (Y1) at (0, 0) {$Y$};
	\node (X2) at (6, 0) {$X$}; \node (Y2) at (6, 6) {$Y$};
	\node [circle, draw] (f1) at (0, 3) {$f$}; \node [circle, draw] (f2) at (6, 3) {$f$};
	\draw [->] (X1) to (f1); \draw [->] (f1) to (Y1);
	\draw [->] (X2) to (f2); \draw [->] (f2) to (Y2);
}
\caption{Covariant and Contravariant Computations}
\label{fig:cocontra}
\end{figure}

The second family of components are the \emph{computations}.
A computation is defined by a function $f : X \to Y$, which can be considered as an open game either \emph{covariantly} as $f : X \to Y$, or \emph{contravariantly} as $f^* : Y^* \to X^*$ (again, this terminology and notation is from linear algebra).
The string diagrams corresponding to the computations are shown in Figure \ref{fig:cocontra}.

\begin{figure}[H]
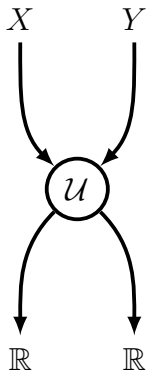

\stringdiagram{
	\node (X) at (0, 6) {$X$}; \node (Y) at (2, 6) {$Y$}; \node (R1) at (0, 0) {$\R$}; \node (R2) at (2, 0) {$\R$};
	\node [circle, draw] (U) at (1, 3) {$\mathcal U$};
	\draw [->] (X) to [out=-90, in=135] (U); \draw [->] (Y) to [out=-90, in=45] (U);
	\draw [->] (U) to [out=-135, in=90] (R1); \draw [->] (U) to [out=-45, in=90] (R2);
}
\caption{Utility function}
\label{fig:utility-function}
\end{figure}

If a function has multiple inputs or outputs then they are denoted by multiple strings.
For example, a utility function $\mathcal U : X \times Y \to \R^2$ can be seen as an open game with type $\mathcal U : X \otimes Y \to \R \otimes \R$, with the string diagram shown in Figure \ref{fig:utility-function}.

\subsubsection{Identities}

\begin{figure}[H]
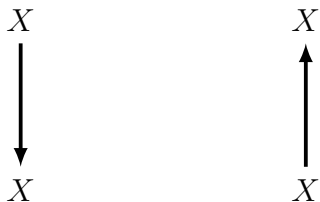

\stringdiagram{
	\node (X1) at (0, 3) {$X$}; \node (X2) at (0, 0) {$X$}; \node (X3) at (5, 3) {$X$}; \node (X4) at (5, 0) {$X$};
	\draw [->] (X1) to (X2); \draw [->] (X4) to (X3);
}
\caption{Identities}
\label{fig:identities}
\end{figure}

Certain functions are given a special notation as computations.
First, we consider the identity function $\id_X : X \to X$, which represents doing nothing.
This is denoted only by a string, without any node.
The computation $\id_X : X \to X$ and its contravariant counterpart $\id_X^* : X^* \to X^*$ are shown in Figure \ref{fig:identities}.

\subsubsection{Deleting and Copying}

\begin{figure}[H]
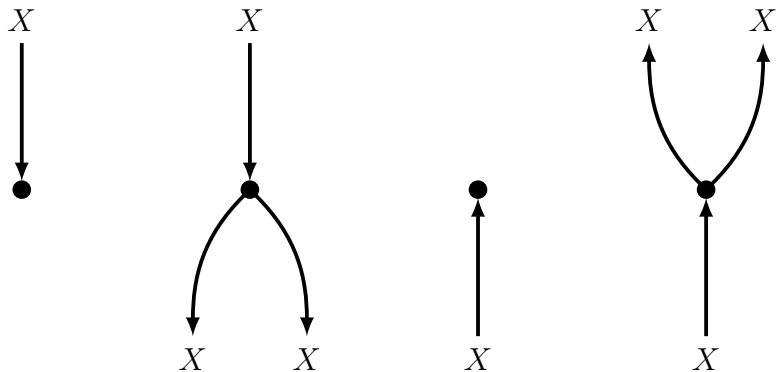

\stringdiagram{
	\node (X1) at (0, 6) {$X$}; \node [circle, scale=0.5, fill=black, draw] (e1) at (0, 3) {};
	\draw [->] (X1) to (e1);
	\node (X2) at (4, 6) {$X$}; \node [circle, scale=0.5, fill=black, draw] (m1) at (4, 3) {};
	\node (X3) at (3, 0) {$X$}; \node (X4) at (5, 0) {$X$};
	\draw [->] (X2) to (m1); \draw [->] (m1) to [out=-135, in=90] (X3); \draw [->] (m1) to [out=-45, in=90] (X4);
	\node [circle, scale=0.5, fill=black, draw] (e2) at (8, 3) {}; \node (X5) at (8, 0) {$X$};
	\draw [->] (X5) to (e2);
	\node (X6) at (11, 6) {$X$}; \node (X7) at (13, 6) {$X$};
	\node [circle, scale=0.5, fill=black, draw] (m2) at (12, 3) {}; \node (X8) at (12, 0) {$X$};
	\draw [->] (X8) to (m2); \draw [->] (m2) to [out=135, in=-90] (X6); \draw [->] (m2) to [out=45, in=-90] (X7);
}
\caption{Cocommutative comonoid and commutative monoid operations}
\label{fig:alg-coalg}
\end{figure}

Next, we consider the deleting function $\operatorname{!}_X : X \to I$, $\operatorname{!}_X (x) = \bullet$ that erases a piece of information, and the copying function $\Delta_X : X \to X \times X$, $\Delta_X (x) = (x, x)$.
Each of these can be covariant or contravariant, giving us four different computations
\begin{eqnarray*}
&\operatorname{!}_X : X \to I \qquad\qquad &\Delta_X : X \to X \otimes X \\
&\operatorname{!}_X^* : I \to X^* \qquad\qquad &\Delta_X^* : X^* \to X^* \otimes X^*
\end{eqnarray*}
The pair $(\operatorname{!}_X, \Delta_X)$ is called a \emph{cocommutative comonoid} on $X$, and the pair $(\operatorname{!}_X^*, \Delta_X^*)$ is called a \emph{commutative monoid} on $X^*$.
The traditional string-diagram notation for these computations is shown in Figure \ref{fig:alg-coalg}.

\subsubsection{Constants}

\begin{figure}[H]
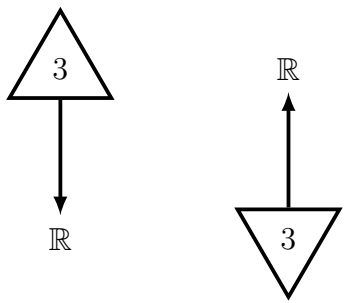

\stringdiagram{
	\node [triangle, draw] (G1) at (1, 3) {$3$};
	\node (R1) at (1, 0) {$\R$};
	\draw [->] (G1) to (R1);
	\node [inverted triangle, draw] (G2) at (5, 0) {$3$};
	\node (R2) at (5, 3) {$\R$};
	\draw [->] (G2) to (R2);
}
\caption{Constants}
\label{fig:constants}
\end{figure}

A constant element of a set, for example $3 \in \R$, is represented by the function $I \to \R$ given by $\bullet \mapsto 3$.
This allows us to consider covariant and contravariant computations $3 : I \to \R$ and $3^* : \R^* \to I$, which are shown in Figure \ref{fig:constants}.

\subsubsection{Braiding}

\begin{figure}[H]
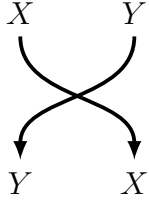

\stringdiagram{
	\node (X1) at (0, 3) {$X$}; \node (Y1) at (2, 3) {$Y$}; \node (X2) at (2, 0) {$X$}; \node (Y2) at (0, 0) {$Y$};
	\draw [->] (X1) to [out=-90, in=90] (X2); \draw [->] (Y1) to [out=-90, in=90] (Y2);
}
\caption{Braiding}
\label{fig:braiding}
\end{figure}

Another computation with a special diagrammatic notation is \emph{braiding} $\sigma_{X, Y} : X \otimes Y \to Y \otimes X$, given by the function $\sigma_{X, Y} (x, y) = (y, x)$.
As the name suggests, this is denoted by a braiding of strings, as shown in Figure \ref{fig:braiding}.
More generally, for any lists of sets that are permutations of each other, such as $A \otimes B^* \otimes C^* \otimes D$ and $C^* \otimes B^* \otimes D \otimes A$, there is a braiding computation that takes one to the other.

\subsubsection{Counit}

The final atomic open games we need to introduce are called the \emph{counits}.
For each set $X$ there is a counit $\tau_X : X \otimes X^* \to I$, which is denoted the bent string in Figure \ref{fig:counit}.
\begin{figure}[H]
\stringdiagram{
	\node (X1) at (2, 2) {$X$}; \node (X2) at (0, 2) {$X$};
	\draw [->] (X1) to [out=-90, in=0] (1, 0) to [out=180, in=-90] (X2);
}
\caption{Counit}
\label{fig:counit}
\end{figure}

Crucially, a string that is bent upwards \emph{does not} represent a well-formed open game, and this is the sole difference between string diagrams in game theory and string diagrams in linear algebra and quantum mechanics.

\subsection{Composition Operations}\label{sec:composition_operations}

\begin{figure}[H]
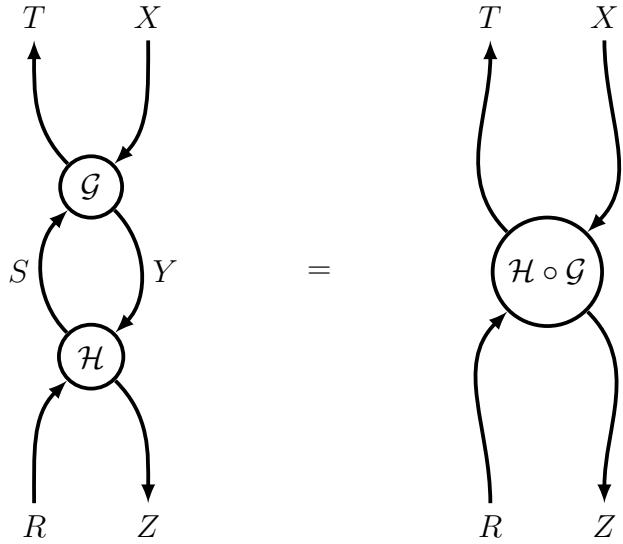

\stringdiagram{
	\node (X1) at (2, 9) {$X$}; \node (Z1) at (2, 0) {$Z$};
	\node (R1) at (0, 0) {$R$}; \node (T1) at (0, 9) {$T$};
	\node [circle, draw] (G) at (1, 6) {$\G$}; \node [circle, draw] (H) at (1, 3) {$\H$};
	\draw [->] (X1) to [out=-90, in=45] (G); \draw [->] (G) to [out=135, in=-90] (T1);
	\draw [->] (G) to [out=-45, in=45] node [right] {$Y$} (H);
	\draw [->] (H) to [out=135, in=-135] node [left] {$S$} (G);
	\draw [->] (H) to [out=-45, in=90] (Z1); \draw [->] (R1) to [out=90, in=-135] (H);
	\node (e) at (5, 4.5) {$=$};
	\node (X2) at (10, 9) {$X$}; \node (Z2) at (10, 0) {$Z$};
	\node (R2) at (8, 0) {$R$}; \node (T2) at (8, 9) {$T$};
	\node [circle, draw] (HoG) at (9, 4.5) {$\H \circ \G$};
	\draw [->] (X2) to [out=-90, in=45] (HoG); \draw [->] (HoG) to [out=-45, in=90] (Z2);
	\draw [->] (R2) to [out=90, in=-135] (HoG); \draw [->] (HoG) to [out=135, in=-90] (T2);
}
\caption{Categorical Composition}
\label{fig:catcomp}
\end{figure}

\begin{figure}[H]
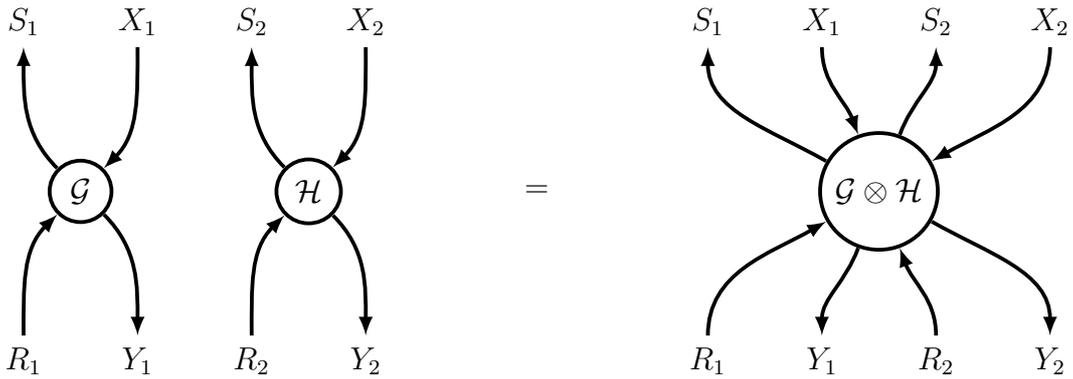

\stringdiagram{
	\node (X11) at (2, 6) {$X_1$}; \node (Y11) at (2, 0) {$Y_1$};
	\node (R11) at (0, 0) {$R_1$}; \node (S11) at (0, 6) {$S_1$};
	\node [circle, draw] (G) at (1, 3) {$\G$};
	\draw [->] (X11) to [out=-90, in=45] (G); \draw [->] (G) to [out=-45, in=90] (Y11);
	\draw [->] (R11) to [out=90, in=-135] (G); \draw [->] (G) to [out=135, in=-90] (S11);
	\node (X21) at (6, 6) {$X_2$}; \node (Y21) at (6, 0) {$Y_2$};
	\node (R21) at (4, 0) {$R_2$}; \node (S21) at (4, 6) {$S_2$};
	\node [circle, draw] (H) at (5, 3) {$\H$};
	\draw [->] (X21) to [out=-90, in=45] (H); \draw [->] (H) to [out=-45, in=90] (Y21);
	\draw [->] (R21) to [out=90, in=-135] (H); \draw [->] (H) to [out=135, in=-90] (S21);
	\node (e) at (9, 3) {$=$};
	\node (X12) at (14, 6) {$X_1$}; \node (Y12) at (14, 0) {$Y_1$};
	\node (R12) at (12, 0) {$R_1$}; \node (S12) at (12, 6) {$S_1$};
	\node (X22) at (18, 6) {$X_2$}; \node (Y22) at (18, 0) {$Y_2$};
	\node (R22) at (16, 0) {$R_2$}; \node (S22) at (16, 6) {$S_2$};
	\node [circle, draw] (GxH) at (15, 3) {$\G \otimes \H$};
	\draw [->] (X12) to [out=-90, in=110] (GxH); \draw [->] (GxH) to [out=-110, in=90] (Y12);
	\draw [->] (R12) to [out=90, in=-150] (GxH); \draw [->] (GxH) to [out=150, in=-90] (S12);
	\draw [->] (X22) to [out=-90, in=30] (GxH); \draw [->] (GxH) to [out=-30, in=90] (Y22);
	\draw [->] (R22) to [out=90, in=-70] (GxH); \draw [->] (GxH) to [out=70, in=-90] (S22);
}
\caption{Categorical Tensor Product}
\label{fig:cattensor}
\end{figure}

There are two basic operations that are used to combine string diagrams together:
\begin{enumerate}
\item \emph{Categorical composition} is a primitive form of sequential play, which in the graphical language is end-to-end joining of strings, shown in Figure \ref{fig:catcomp};
\item \emph{Tensor product} is a primitive form of simultaneous play, which in the graphical language is side-by-side juxtaposition, shown in Figure \ref{fig:cattensor}.
\end{enumerate}

In the setting of linear algebra, categorical composition corresponds to matrix multiplication (or, equivalently, composition of linear operators), and tensor product corresponds to tensor product of matrices or linear operators.

\begin{figure}[H]
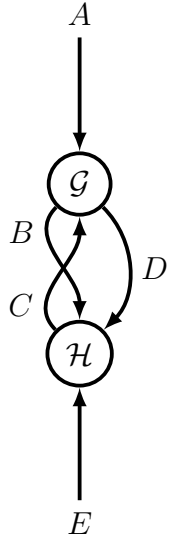

\stringdiagram{
	\node (A) at (1, 9) {$A$}; \node (E) at (1, 0) {$E$};
	\node [circle, draw] (G) at (1, 6) {$\G$}; \node [circle, draw] (H) at (1, 3) {$\H$};
	\draw [->] (A) to (G); \draw [->] (E) to (H);
	\draw [->] (G) to [out=-45, in=45] node [right] {$D$} (H);
	\draw [->] (G) to [out=-135, in=90] node [left, near start] {$B$} (H);
	\draw [->] (H) to [out=135, in=-90] node [left, near start] {$C$} (G);
}
\caption{Categorical Composition with Braiding}
\label{fig:composition-example}
\end{figure}

In general, two open games can be composed sequentially only when the strings and their labels (which are sets) exactly match, just like when we compose functions the domain of one must match the codomain of the other.
If they differ only up to a permutation, they can be made composable by first composing with a suitable braiding.
For example, if we have a game $\G : A \to B \otimes C^* \otimes D$ and $\H : C^* \otimes B \otimes D \to E^*$ then we can form the composition
\[ \H \circ \sigma \circ \G : A \to E^* \]
where $\sigma : B \otimes C^* \otimes D \to C^* \otimes B \otimes D$ is a braiding, which is shown in Figure \ref{fig:composition-example}.
On the other hand, we may form the tensor product of any two games, with no restrictions on the matching of types.

Finally, we come to the most crucial point: the things which we can compose need not be represented by only a node (which is used to represent an `arbitrary' open game), but can themselves be complicated aggregates formed by categorical composition and tensor product.
We need only keep track of the strings entering and leaving the top and bottom of the diagrams.
Composition, in general, is pasting entire diagrams top-to-bottom and joining the string together, and tensor product in general is pasting entire diagrams side-by-side.

The data $(\Sigma, \P, \C, \E)$ for both categorical compositions and tensor products is computed in a uniform way from the same data for the two diagrams being composed.
The appropriate definitions are given in the appendix.

\subsection{Algebraic Substitution}
\label{sec:substitution}

In this section we will show the power of algebraic compositionality and its visual representation for the canonical open games.
The graphical device we introduce is what we call \emph{boxing}.
It is the diagrammatic equivalent of the algebraic process of naming subexpressions in equations,
like in $E = (a + b )x$ which is turned into $E = S x$ where $S = (a + b)$.
Specifically we draw a box around some complicated open game and forget about its internal structure and only use its connections to the outside.
The box can then be reused in different places.
Boxing fosters efficiency in economic model building and helps to cope with complex models.

Boxing is strongly analogous to the creation of interfaces, classes and APIs in software engineering.
In particular, there is a certain art to designing good interfaces that decrease programmer workload by facilitating code reuse and useful abstractions, 
as opposed to bad interfaces which increase programmer workload by adding meaningless complexity, boilerplate code and unhelpful abstractions.

\begin{figure}[H]
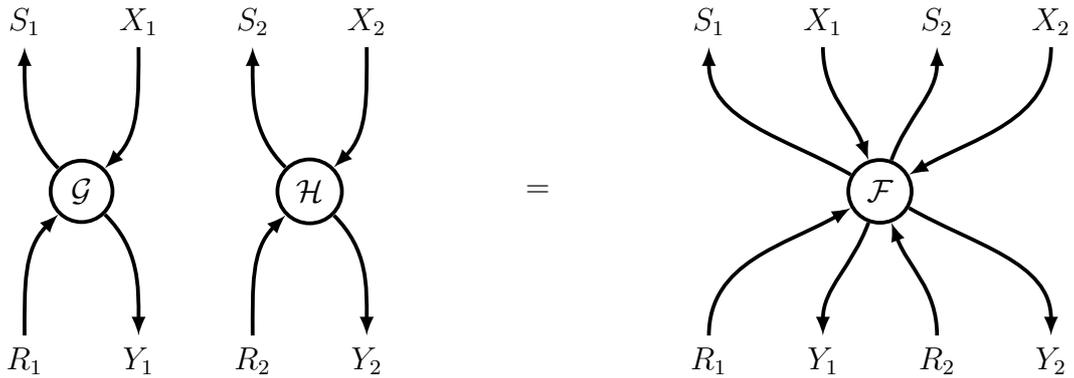

\stringdiagram{
	\node (X11) at (2, 6) {$X_1$}; \node (Y11) at (2, 0) {$Y_1$};
	\node (R11) at (0, 0) {$R_1$}; \node (S11) at (0, 6) {$S_1$};
	\node [circle, draw] (G) at (1, 3) {$\G$};
	\draw [->] (X11) to [out=-90, in=45] (G); \draw [->] (G) to [out=-45, in=90] (Y11);
	\draw [->] (R11) to [out=90, in=-135] (G); \draw [->] (G) to [out=135, in=-90] (S11);
	\node (X21) at (6, 6) {$X_2$}; \node (Y21) at (6, 0) {$Y_2$};
	\node (R21) at (4, 0) {$R_2$}; \node (S21) at (4, 6) {$S_2$};
	\node [circle, draw] (H) at (5, 3) {$\H$};
	\draw [->] (X21) to [out=-90, in=45] (H); \draw [->] (H) to [out=-45, in=90] (Y21);
	\draw [->] (R21) to [out=90, in=-135] (H); \draw [->] (H) to [out=135, in=-90] (S21);
	\node (e) at (9, 3) {$=$};
	\node (X12) at (14, 6) {$X_1$}; \node (Y12) at (14, 0) {$Y_1$};
	\node (R12) at (12, 0) {$R_1$}; \node (S12) at (12, 6) {$S_1$};
	\node (X22) at (18, 6) {$X_2$}; \node (Y22) at (18, 0) {$Y_2$};
	\node (R22) at (16, 0) {$R_2$}; \node (S22) at (16, 6) {$S_2$};
	\node [circle, draw] (GxH) at (15, 3) {$\mathcal{F}$};
	\draw [->] (X12) to [out=-90, in=110] (GxH); \draw [->] (GxH) to [out=-110, in=90] (Y12);
	\draw [->] (R12) to [out=90, in=-150] (GxH); \draw [->] (GxH) to [out=150, in=-90] (S12);
	\draw [->] (X22) to [out=-90, in=30] (GxH); \draw [->] (GxH) to [out=-30, in=90] (Y22);
	\draw [->] (R22) to [out=90, in=-70] (GxH); \draw [->] (GxH) to [out=70, in=-90] (S22);
}
\caption{Box of the Categorical Tensor Product}
\label{fig:cattensor_boxing}
\end{figure}

Substituting equals for equals where one renames composed structures is closely related to the sequential and parallel compositions in Figure \ref{fig:catcomp} and Figure \ref{fig:cattensor}.
\begin{figure}[H]
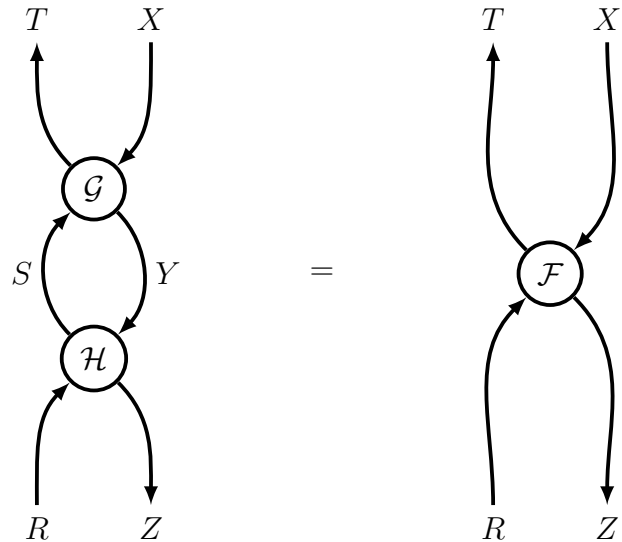

\stringdiagram{
	\node (X1) at (2, 9) {$X$}; \node (Z1) at (2, 0) {$Z$};
	\node (R1) at (0, 0) {$R$}; \node (T1) at (0, 9) {$T$};
	\node [circle, draw] (G) at (1, 6) {$\G$}; \node [circle, draw] (H) at (1, 3) {$\H$};
	\draw [->] (X1) to [out=-90, in=45] (G); \draw [->] (G) to [out=135, in=-90] (T1);
	\draw [->] (G) to [out=-45, in=45] node [right] {$Y$} (H);
	\draw [->] (H) to [out=135, in=-135] node [left] {$S$} (G);
	\draw [->] (H) to [out=-45, in=90] (Z1); \draw [->] (R1) to [out=90, in=-135] (H);
	\node (e) at (5, 4.5) {$=$};
	\node (X2) at (10, 9) {$X$}; \node (Z2) at (10, 0) {$Z$};
	\node (R2) at (8, 0) {$R$}; \node (T2) at (8, 9) {$T$};
	\node [circle, draw] (HoG) at (9, 4.5) {$\mathcal{F}$};
	\draw [->] (X2) to [out=-90, in=45] (HoG); \draw [->] (HoG) to [out=-45, in=90] (Z2);
	\draw [->] (R2) to [out=90, in=-135] (HoG); \draw [->] (HoG) to [out=135, in=-90] (T2);
}
\caption{Box of the Categorical Composition}
\label{fig:catcomp_boxing}
\end{figure}
In Figure \ref{fig:catcomp_boxing} we see the substitution or renaming with respect to the sequential composition where the composed structure is simple renamed $\mathcal{F} = \H \circ \G$
In the same way substitution works for the parallel composition in Figure \ref{fig:cattensor_boxing} where $\mathcal{F}=\G \otimes \H$.
In the algebraic notation renaming subexpressions is trivial but as a graphical device it is essential for coping with the complexity involved in complicated games.

\subsection{Topological Moves}

As mentioned in section \label{sec:background}, an important feature of string diagrams is that they are \emph{invariant} under topological moves.
That is, if we deform the string diagram, the meaning does not change. Formally, it is a theorem that the data $(\Sigma, \P, \C, \E)$ is invariant under topological moves; these topological moves are related to Reidemeister moves in knot theory.

\begin{figure}[H]
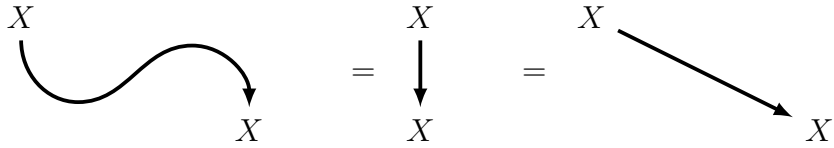

\stringdiagram{
	\node (X1) at (0, 3) {$X$};
	\node (X2) at (4, 1) {$X$};
	\node at (6,2) {$=$};
	\node (X3) at (7, 3) {$X$};
	\node (X4) at (7, 1) {$X$};
	\node at (9,2) {$=$};
	\node (X5) at (10, 3) {$X$};
	\node (X6) at (14, 1) {$X$};
	\draw [->] (X1) to [out=-90,in=180] (1,1.5) to [out=0,in=180] (3,2.5) to [out=0,in=90] (X2);
	\draw [->] (X3) to (X4);
	\draw [->] (X5) to (X6);
}
\caption{Yanking in a Compact Closed Category}
\label{fig:yanking}
\end{figure}

We need to be a bit more formal about exactly what constitutes a topological move.
The paper \cite{selinger11} gives a survey of many variants of the string diagram language for various different axiomatic settings arising in various application areas.
The situation in game theory falls part way between two of the best known:
\begin{itemize}
\item \emph{Symmetric monoidal categories} allow arrows pointing only downwards, and are invariant under topological moves that preserve this. In particular, an arrow can never be `bent around' to point in the opposite direction.
\item \emph{Compact closed categories} allow strings pointing in both directions, and allow both counits (bending a downward-pointing string to point upwards) and units (bending an upward-pointing string to point downwards). There are no restrictions on the invariant topological moves. In particular, compact closed categories satisfy the \emph{yanking equation} allowing a bent string to be straightened, illustrated in Figure \ref{fig:yanking}.
\end{itemize}

In game theory the unit is not a well-formed string diagram, as previously mentioned.
Therefore we have one restriction on the invariant moves: two diagrams denote the same game if one can be deformed into the other without creating a unit, or downward-bend.

\begin{figure}[H]
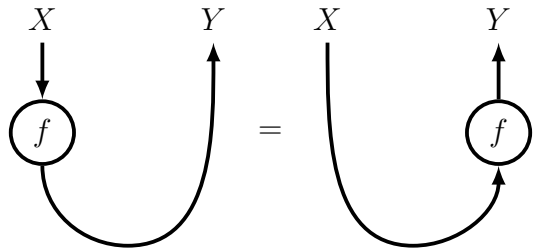

\stringdiagram{
	\node (X1) at (0,4) {$X$};
	\node [circle,draw] (f1) at (0,2) {$f$};
	\node (Y1) at (3,4) {$Y$};
	\draw [->] (X1) to [out=-90,in=90] (f1);
	\draw [->] (f1) to [out=-90,in=180] (1.5,0) to [out=0,in=-90] (Y1);

	\node at (4,2) {$=$};

	\node (X2) at (5,4) {$X$};
	\node [circle,draw] (f2) at (8,2) {$f$};
	\node (Y2) at (8,4) {$Y$};
	\draw [->] (X2) to [out=-90,in=180] (6.5,0) to [out=0,in=-90] (f2);
	\draw [->] (f2) to [out=90,in=-90] (Y2);
}
\caption{Counit Law}
\label{fig:counit-law}
\end{figure}

An example of a move we are allowed to make is shown in Figure \ref{fig:counit-law}, which slides a computation around a counit.
This is called the \emph{counit law} and is also true in a compact closed category.

\begin{figure}[H]
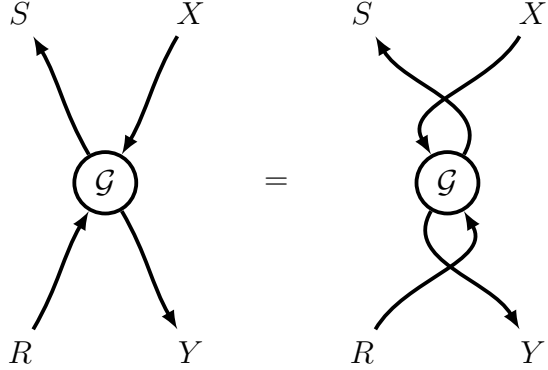

\stringdiagram{
	\node (S1) at (0,6){$S$};
	\node (X1) at (3,6){$X$};
	\node (R1) at (0,0){$R$};
	\node (Y1) at (3,0){$Y$};
	\node [circle,draw] (G1) at (1.5,3){$\mathcal{G}$};

	\draw [->] (G1) to [out=120,in=-60] (S1);
	\draw [->] (X1) to [out=240,in=60] (G1);
	\draw [->] (R1) to [out=60,in=240] (G1);
	\draw [->] (G1) to [out=-60,in=120] (Y1);

	\node (e) at (4.5,3){$=$};

	\node (S2) at (6,6){$S$};
	\node (X2) at (9,6){$X$};
	\node (R2) at (6,0){$R$};
	\node (Y2) at (9,0){$Y$};
	\node [circle,draw] (G2) at (7.5,3){$\mathcal{G}$};

	\draw [->] (G2) to [out=60,in=-60] (S2);
	\draw [->] (X2) to [out=240,in=120] (G2);
	\draw [->] (R2) to [out=60,in=-60] (G2);
	\draw [->] (G2) to [out=240,in=120] (Y2);
}
\caption{Port Braiding}
\label{fig:port_braiding}
\end{figure}

We are allowing to relabel the ports in order to keep string diagrams symmetric where we have for example several parallel open games,
see Figure \ref{fig:port_braiding}.

\subsection{Closed Games}
\begin{definition}[Closed Game]
A \emph{closed game} is a game of the form $\G : I \to I$, which has $X = Y = R = S$ all containing a single element $\bullet$. Closed games have trivial play and coplay functions (because they can only return $\bullet$), and their equilibrium function reduces to a unary relation $\E_\G \subseteq \Sigma_\G$ (because $X$ and $Y \to R$ both contain only a single element). If $\sigma$ is a strategy profile for a closed game $\G$ with $\sigma \in \E_\G$, we will call $\sigma$ an \emph{equilibrium} of $\G$.
\end{definition}

We will show by example that many games in the ordinary sense can be described as closed games, and that this equilibrium condition agrees with the ordinary Nash equilibrium. Nevertheless the theory necessarily focusses on general games, which are theoretically better behaved. The idea is that we start with building blocks that are non-closed or open, and compose them with operations that cannot be defined only on closed games, but in the end we obtain a game that is closed and describes something that, intuitively speaking, is a game.

However, games are a very overloaded notion and we want call a game in this paper what is actually a closed game in our definition.

\section{Examples}\label{sec:games}
In this section we provide examples of decisions and games represented by string diagrams 
along with the algebraic expressions of the string diagrams and the data of open games.

Secondly, we give examples where we show applications of the composition of more complex games from basic building blocks.
We show the programatic technique of boxing.
Boxing hides the interior of some complicated open game into a box and then uses this box for a simplified composition using the operators that we have defined.
This is useful for representing the hierarchy of games or simply for the purpose of reusing common parts in several games.

\subsection{Decisions}\label{sec:decisions}
\begin{figure}[H]
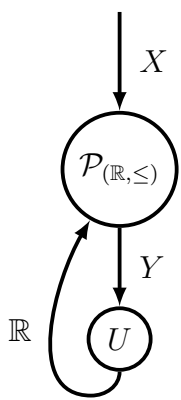

\stringdiagram{
	\node (X3) at (8,9){};
	\node [circle, draw] (P3) at (8,6){$\mathcal{P}_{(\R,\leq)}$};
	\node [circle, draw] (q3) at (8,3){$U$};
	\draw [->] (X3) to [out = -90, in = 90] node [right = 0.1cm, midway] {$X$} (P3);
	\draw [->] (P3) to [out = -90, in = 90] node [right = 0.1cm, midway] {$Y$} (q3);
	\draw [->] (q3) to [out = -90, in = 0] (7.5,2) to [out=180,in=240] node [left = 0.1cm, midway] {$\R$} (P3);
}
\caption{Simple Decision Problem}
\label{fig:dec}
\end{figure}

Our framework allows to uniformly treat decision theory and game theory 
such that games are composed from decisions, see \cite{Hedges_et_al_2015_games}.
Similarly a game can be seen from the perspective of an agent to be a decision problem where the actions of the other agents is subsumed into the context of the deciding agent.
However, a decision is a game with only one player and who is making one move.
The simplest possible decision problem is depicted in Figure \ref{fig:dec}. 
A player observes the history $X$ and makes a decision $\mathcal{P}_{(\R,\leq)}$ with the action being of type $Y$. 
He receives a utility $U$ of type  $\mathbb R$ from his decision that is taking into account the order relation $\leq$ on the utility type.

The decision is given in algebraic notation as 
\[
\mathcal G = \tau_{\mathbb R} \circ (\mathcal U \otimes \mathbb R^*) \circ P : X \to  I
\]
where the data $\mathcal G = (\Sigma_\mathcal G, \mathbf P_\mathcal G, \mathbf C_\mathcal G, \mathbf E_\mathcal G)$ specifying $\mathcal G$ is given by
\begin{itemize}
	\item $\Sigma_\mathcal G = X \to Y$
	\item $\mathbf P_\mathcal G : \Sigma_\mathcal G  \times X \to 1$, $\mathbf P_\mathcal G (\sigma, x) = \bullet$
	\item $\mathbf C_\mathcal G : \Sigma_\mathcal G \times X \times 1 \to 1$, $\mathbf C_\mathcal G (\sigma, x, \bullet) = \bullet$
	\item $\mathbf E_\mathcal G : X \times (1 \to 1) \to \mathcal P \Sigma_\mathcal G$, $\mathbf E_\mathcal G (x, k) = \{ \sigma : X \to Y \mid k (\sigma x) = \max k \}$
\end{itemize}

In this paper we focus on agents whose behavior is described by utility maximization.
However, in our framework it is equally possible to model agents who are choosing directly according to some preference relation without the detour via a utility function.
We can formally take this structure into account in its graphical representation.

\begin{figure}[H]
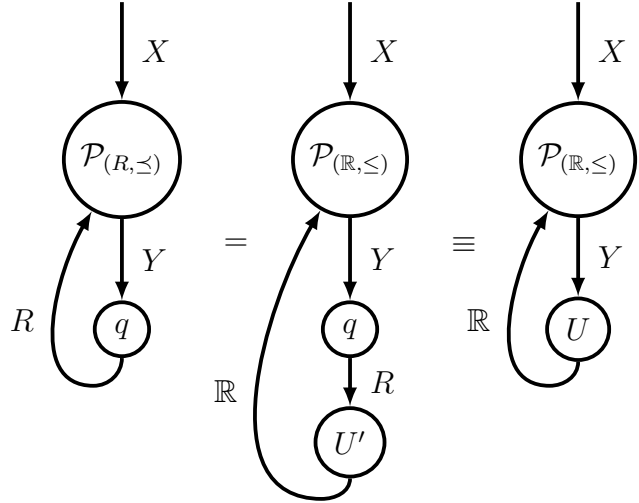

\stringdiagram{
	\node (X1) at (0,9){};
	\node [circle, draw] (P1) at (0,6){$\mathcal{P}_{(R,\preceq)}$};
	\node [circle, draw] (q1) at (0,3){$q$};
	\draw [->] (X1) to [out = -90, in = 90] node [right = 0.1cm, midway] {$X$} (P1);
	\draw [->] (P1) to [out = -90, in = 90] node [right = 0.1cm, midway] {$Y$} (q1);
	\draw [->] (q1) to [out = -90, in = 0] (-0.5,2) to [out=180,in=240] node [left = 0.1cm, midway] {$R$} (P1);

	\node (e1) at (2,4.5){$=$};

	\node (X2) at (4,9){};
	\node [circle, draw] (P2) at (4,6){$\mathcal{P}_{(\R,\leq)}$};
	\node [circle, draw] (q2) at (4,3){$q$};
	\node [circle, draw] (i2) at (4,1){$U'$};
	\draw [->] (X2) to [out = -90, in = 90] node [right = 0.1cm, midway] {$X$} (P2);
	\draw [->] (P2) to [out = -90, in = 90] node [right = 0.1cm, midway] {$Y$} (q2);
	\draw [->] (q2) to [out = -90, in = 90] node [right = 0.1cm, midway] {$R$} (i2);
	\draw [->] (i2) to [out = -90, in = 0] (3.5,0) to [out=180,in=240] node [left = 0.1cm, midway] {$\R$} (P2);

	\node (e2) at (6,4.5){$\equiv$};

	\node (X3) at (8,9){};
	\node [circle, draw] (P3) at (8,6){$\mathcal{P}_{(\R,\leq)}$};
	\node [circle, draw] (q3) at (8,3){$U$};
	\draw [->] (X3) to [out = -90, in = 90] node [right = 0.1cm, midway] {$X$} (P3);
	\draw [->] (P3) to [out = -90, in = 90] node [right = 0.1cm, midway] {$Y$} (q3);
	\draw [->] (q3) to [out = -90, in = 0] (7.5,2) to [out=180,in=240] node [left = 0.1cm, midway] {$\R$} (P3);
}
\caption{Preference and Utility Based Decisions}
\label{fig:dec:prefs}
\end{figure}

In Figure \ref{fig:dec:prefs} we see on the left hand side a move $Y$ that is the argument of an outcome function $q$ with the result type $R$.
The result is taken into consideration in the decision $\mathcal{P}_{(R,\preceq)}$ based on some order $\preceq$ which may be any preference relation.
In the middle string diagram we describe how to turn the preference order into utility functions. 
Here the results $R$ of the outcome function $q$ is translated by the utility function $U'$ into a real number $\mathbb R$.
The decision $\mathcal{P}_{(\R,\leq)}$ is then based on the usual $\leq$ ordering on $\mathbb R$.
Equivalently, we see on the right hand side the string diagram where the utility is defined directly on the action of type $Y$.

Algebraically, these equations are
\begin{align*}
	&\tau_R \circ (R^* \otimes q) \circ \mathcal P_R \\
	=\ &\tau_\R \circ (\R^* \otimes (U' \circ q)) \circ \mathcal P_\R \\
	\equiv\ &\tau_\R \circ (\R \otimes U) \circ \mathcal P_\R
\end{align*}
The second equality is simply a definitional equality $U \equiv U' \circ q$, but the first equality arises if the preference relation can be represented by a  utility function, which is the case under some standard conditions, 
see \cite{Fudenberg1991,Mas-Colell1995}.

\subsection{Parallel Composition}\label{sec:parallel}
We now want to discuss the first of the important composition operators (combinators) that take open games as arguments and return more complex ones.
There are two combinators of this kind in this paper. 
A parallel and a sequential combination of open games, meaning that the games are played in parallel or sequentially.

\subsubsection{Bimatrix games}
\begin{figure}[H]
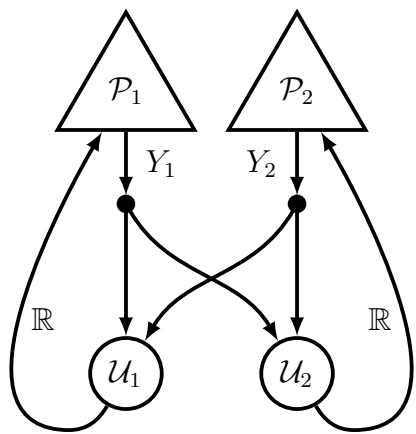

\stringdiagram{
	\node [triangle, draw] (P1) at (0,6) {$\mathcal{P}_1$};
	\node [triangle, draw] (P2) at (3,6) {$\mathcal{P}_2$};
	\node [circle, draw] (U1) at (0,1) {$\mathcal{U}_1$};
	\node [circle, draw] (U2) at (3,1) {$\mathcal{U}_2$};
	\node [circle, scale=0.5, fill=black, draw] (c1) at (0, 4) {};
	\node [circle, scale=0.5, fill=black, draw] (c2) at (3, 4) {};
	\draw [->] (P1) to [out = -90, in = 90] node [right = 0.1cm, midway] {$Y_1$} (c1);
	\draw [->] (c1) to [out = -90, in = 90] (U1);
	\draw [->] (c1) to [out = -60, in = 120] (U2);	
	\draw [->] (P2) to [out = -90, in = 90] node [left = 0.1cm, midway] {$Y_2$} (c2);
	\draw [->] (c2) to [out = 240, in = 60] (U1);
	\draw [->] (c2) to [out = -90, in = 90] (U2);
	\draw [->] (U1) to [out=-120, in=0] (-1,0) to [out=180,in=240] node [right] {$\R$} (P1);
	\draw [->] (U2) to [out=-60, in=180] (4,0) to [out=0,in=-60] node [left] {$\R$} (P2);
}
\caption{Bimatrix Game}
\label{fig:batt}
\end{figure}

Firstly we consider a classical bimatrix game where two players maximize their utility functions defined over the reals $\mathbb R$. 
We denote the set of the first player's actions with $Y_1$ and the actions of the second player with $Y_2$. 
This game can be represented by the string diagram in Figure \ref{fig:batt} where $\mathcal{P}_1$ and $\mathcal{P}_2$ depict the decisions of the two players, respectively.  
Both players maximize their utilities $\mathcal{U}_1$ and  $\mathcal{U}_2$ which map the players' actions $Y_1$ and $Y_2$ into an outcome of type $\mathbb R$. 
The utility functions are represented by open game computations.

In this game we see the advantage of using braided ports as allowed in Figure \ref{fig:port_braiding} where without the player $\mathcal{P}_2$ should have received the strings from the utility to the left of the outgoing string of $Y_2$.
Here, braiding allows us to unambiguously twist the utility and the action ports (also the coutility and history ports), and by that highlight the symmetry of the overall game more accurately.
However, the braiding of ports is unambiguous for the outgoing and incoming ports but not for different strings into the same kind of incoming or outgoing port.
Similarly it is not important whether the outgoing strings of the utility nodes are starting at the left (usually defined to be the place for the incoming utility port) or the right (usually defined to be the port for the action of an open game).
For example, in the bimatrix game the two incoming strings of the utility nodes cannot be flipped without changing the algebraic meaning.

In algebraic notation, the closed bimatrix game is given by
\begin{align*}
	\G =\ &(\tau_\R \otimes \tau_R) \circ (\sigma_{\R^*, \R} \otimes \R \otimes \R^*) \circ (\R^* \otimes U \otimes \R^*) \\
	&\circ (\sigma_{Y_1, \R^*} \otimes Y_2 \otimes \R^*) \circ (\mathcal P_1 \otimes \mathcal P_2) : I \to I
\end{align*}
where $U : Y_1 \otimes Y_2 \to \R \otimes \R$ is given by
\[ U = (\mathcal U_1 \otimes \mathcal U_2) \circ (Y_1 \otimes \sigma_{Y_1, Y_2} \otimes Y_2) \circ (\Delta_{Y_1} \otimes \Delta_{Y_2}) \]

Fortunately, this expression can be calculated from the string diagram by a completely automatic procedure, as can its denotation. 
The denotation is a closed game with strategy space $\Sigma_\G = X \times Y$, where the strategy profile $(\sigma_1, \sigma_2)$ is an equilibrium iff
\[ \mathcal U_1 (\sigma_1, \sigma_2) = \max_{y_1 \in Y_1} \mathcal U_1 (y_1, \sigma_2) \]
\[ \mathcal U_2 (\sigma_1, \sigma_2) = \max_{y_2 \in Y_2} \mathcal U_2 (\sigma_1, y_2) \]
which is the usual Nash equilibrium. 

It is instructive to focus on the part of this diagram that contains the players, namely
\[ \mathcal P_1 \otimes \mathcal P_2 : I \to Y_1 \otimes \mathbb R^* \otimes Y_2 \otimes \mathbb R^* \]
The data that specifies this, $(\Sigma_{\mathcal P_1 \otimes \mathcal P_2}, \mathbf P_{\mathcal P_1 \otimes \mathcal P_2}, \mathbf C_{\mathcal P_1 \otimes \mathcal P_2}, \mathbf E_{\mathcal P_1 \otimes \mathcal P_2})$ is given by
\begin{itemize}
	\item $\Sigma_{\mathcal P_1 \otimes \mathcal P_2} = Y_1 \times Y_2$
	\item $\mathbf P_{\mathcal P_1 \otimes \mathcal P_2} : \Sigma_{\mathcal P_1 \otimes \mathcal P_2} \times 1 \to Y_1 \times Y_2$, $\mathbf P_{\mathcal P_1 \otimes \mathcal P_2} ((\sigma_1, \sigma_2), \bullet) = (\sigma_1, \sigma_2)$
	\item $\mathbf C_{\mathcal P_1 \otimes \mathcal P_2} : \Sigma_{\mathcal P_1 \otimes \mathcal P_2} \times 1 \times \mathbb R^2 \to 1$, $\mathbf C_{\mathcal P_1 \otimes \mathcal P_2} ((\sigma_1, \sigma_2), \bullet, (u_1, u_2)) = \bullet$
	\item $\mathbf E_{\mathcal P_1 \otimes \mathcal P_2} : 1 \times (Y_1 \times Y_2 \to \mathbb R^2) \to \mathcal P \Sigma_{\mathcal P_1 \otimes \mathcal P_2}$,
	\[ \mathbf E_{\mathcal P_1 \otimes \mathcal P_2} (\bullet, k) = \left\{ (\sigma_1, \sigma_2) : Y_1 \times Y_2 \ \left|\ \begin{matrix}
		(\pi_1 \circ k) (\sigma_1, \sigma_2) = \max_{y_1 \in Y_1} (\pi_1 \circ k) (y_1, \sigma_2), \\
		(\pi_2 \circ k) (\sigma_1, \sigma_2) = \max_{y_2 \in Y_2} (\pi_2 \circ k) (\sigma_1, y_2)
	\end{matrix} \right. \right\} \]
\end{itemize}
Again, all of this data is derivable by a purely mechanical and computable procedure.

Note, that the string diagram in Figure \ref{fig:batt} represents the information flow in any bimatrix game 
without specifying the actual payoffs, that specializes the general bimatrix game into Prisoner's Dilemma, Meeting in New York or 
any other example that are usually named according to the story motivating the concrete payoff matrices.
However, in specific examples further structures on the payoff matrix, like symmetries, can be explicitly represented 
and visualized in the string diagram, as we do in the next example.

\subsubsection{Visualising symmetric outcomes}
This particular example of a bimatrix game, Meeting in New York, see \cite{Fudenberg1991,Mas-Colell1995}, 
is about two players who have to choose simultaneously one location to meet: either at the 
Grand Central Terminal or the Empire State Building. 
They only receive a positive payoff if they meet in the same place. 
These payoffs are given in Table \ref{tab:NY}.

\begin{table}[ht!]
\begin{center}
\[
\begin{array}{|l|c|c|}\hline
	& GCT             	& ES    \\ \hline
GCT	& 2,2           	& 0,0   \\ \hline
ES  	& 0,0           	& 2,2   \\ \hline
\end{array}
\]
\caption{Meeting in New York}
\label{tab:NY}
\end{center}
\end{table}

\begin{figure}[H]
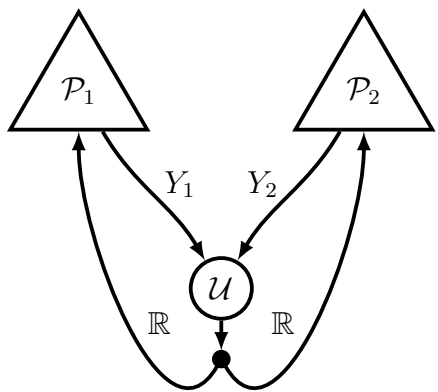

\stringdiagram{
	\node [triangle, draw] (P1) at (1,5) {$\mathcal{P}_1$};
	\node [triangle, draw] (P2) at (6,5) {$\mathcal{P}_2$};
	\node [circle, draw] (U) at (3.5,1.5) {$\mathcal{U}$};
	\node [circle,scale = 0.5, fill=black,draw] (iU) at (3.5,0.25) {};
	\draw [->] (P1) to [out = -60, in = 120] node [above = 0.2cm, near end] {$Y_1$} (U);
	\draw [->] (P2) to [out = -120, in = 60] node [above = 0.2cm, near end] {$Y_2$} (U);
	\draw [->] (U) to [out = -90, in = 90] (iU);
	\draw [->] (iU) to [out=240, in=-90]  node [right = 0.1cm, midway] {$\R$} (P1);
	\draw [->] (iU) to [out=-60, in=-90]  node [left = 0.1cm, midway] {$\R$} (P2);
}
\caption{Meeting in New York}
\label{fig:meet}
\end{figure}

With $Y_1 = Y_2 = \{ \hbox{GCT}, \hbox{ES} \}$ being the action space for both players and 
$\mathcal{U} : Y_1 \times Y_2 \to \R$ given by the bimatrix  in Table \ref{tab:NY}.
This game has a symmetric structure specializing the general bimatrix game.
In the string diagram of Figure \ref{fig:meet}, as opposed to the one in \ref{fig:batt}, 
both players receive utilities from the same function $\mathcal{U}$ 
which maps the actions into the two possible payoffs $0$ or $2$.

In algebraic notation, this closed game is given by
\begin{align*}
\G =\ &(\tau_\R \otimes \tau_\R) \circ (\sigma_{\R^*, \R} \otimes \R \otimes \R^*) \circ (\R^* \otimes (\Delta_\R \otimes \mathcal U) \otimes \R^*) \\
&\circ (\sigma_{Y_1, \R^*} \otimes Y_2 \otimes \R^*) \circ (\mathcal P_1 \otimes \mathcal P_2) : I \to I
\end{align*}

This closed game $\G$ has $\Sigma_\G = Y_1 \times Y_2$, and a strategy profile $(\sigma_1, \sigma_2)$ is an equilibrium iff
\[ \mathcal U (\sigma_1, \sigma_2) = \max_{y_1 \in Y_1} \mathcal U (y_1, \sigma_2) \]
\[ \mathcal U (\sigma_1, \sigma_2) = \max_{y_2 \in Y_2} \mathcal U (\sigma_1, y_2) \]
The equilibrium function $\E$ picks out the pure-strategy Nash equilibria which are for
the particular payoff function $\mathcal U$ given in Table \ref{tab:NY} the strategy profiles $(\text{GCT}, \text{GCT})$ and $(\text{ES}, \text{ES})$.
\subsubsection{Coordination and differentiation games}
\begin{figure}[H]
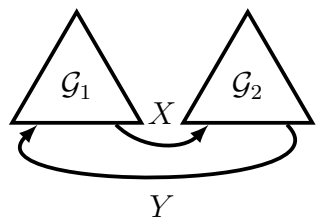

\stringdiagram{
	\node [triangle,draw] (G1) at (1,0){$\mathcal{G}_1$};
	\node [triangle,draw] (G2) at (4,0){$\mathcal{G}_2$};
	\draw [->] (G1) to [out=-45, in=225] node [above = 0.1cm, midway] {$X$} (G2);
	\draw [->] (G2) to [out=-45, in=225] node [below = 0.1cm, midway] {$Y$} (G1);
}
  \caption{Coordination and Differentiation}
\label{fig:2fix}
\end{figure}

The framework of higher-order games, see \cite{Hedges_et_al_2015_decisions,Hedges_et_al_2015_games}, 
can be used to represent behavior not described by maximization but by any operation other than the max operator.
Similarly preferences may not be described by some utility functions or rational preference relations but by any relation.
We only need some structure that picks out some elements of the outcome set.
Specifically, a most interesting example of this kind of higher-order goals can be elegantly described by higher-order functions being a fixed point operator.
The fixed point operator (takes a function as its argument like the max operator) represents goals where the agent wants to coordinate with others players.
This models for example the judges in a Keynesian beauty contest with the goal of voting for the winner.
Note, that here the preferences are not with respect to the candidates or the outcomes per se but dependent of the outcome function and the preferences of the other players, 
the judge with a fixed point goal needs to solve the game first and then decide accordingly.
This fixed point goal in general represents the goal of coordination (or differentiation via an anti-fixed point operator) is given as a string diagram in Figure \ref{fig:2fix} 
where the agents decide upon their actions by taking into account the action of their opponent and not some specific outcome or utility thereof.
Hence, the short circuit in this string diagram is the paradigmatic example of the paradigmatic motive of coordination or differentiation in social systems.

\subsection{Sequential Composition}
In this section, we introduce the string diagram representation of games with sequentially moving players. 

For illustration, consider the ultimatum game \cite{Gueth1982}.
In this game, player 1 offers a split of a pie of size $N$, and player 2 can choose to accept or reject.
Thus the set of moves for player 1 is $Y_1 = \{ 0, 1, \ldots, N \}$ (where the $x \in X$ is interpreted as the offer of $x$ for player 1 and $N - x$ for player 2), and the set of moves for player 2 is $Y_2 = \{ A, R \}$.

The utility functions for the two players are given by the share of the pie  they receive, namely
\[ \mathcal U_1 (y_1, y_2) = \begin{cases} y_1 & \text{ if } y_2 = A \\ 0 & \text{ if } y_2 = R \end{cases} \qquad\qquad \mathcal U_2 (y_1, y_2) = \begin{cases} N - y_1 & \text{ if } y_2 = A \\ 0 & \text{ if } y_2 = R \end{cases} \]

\begin{figure}[H]
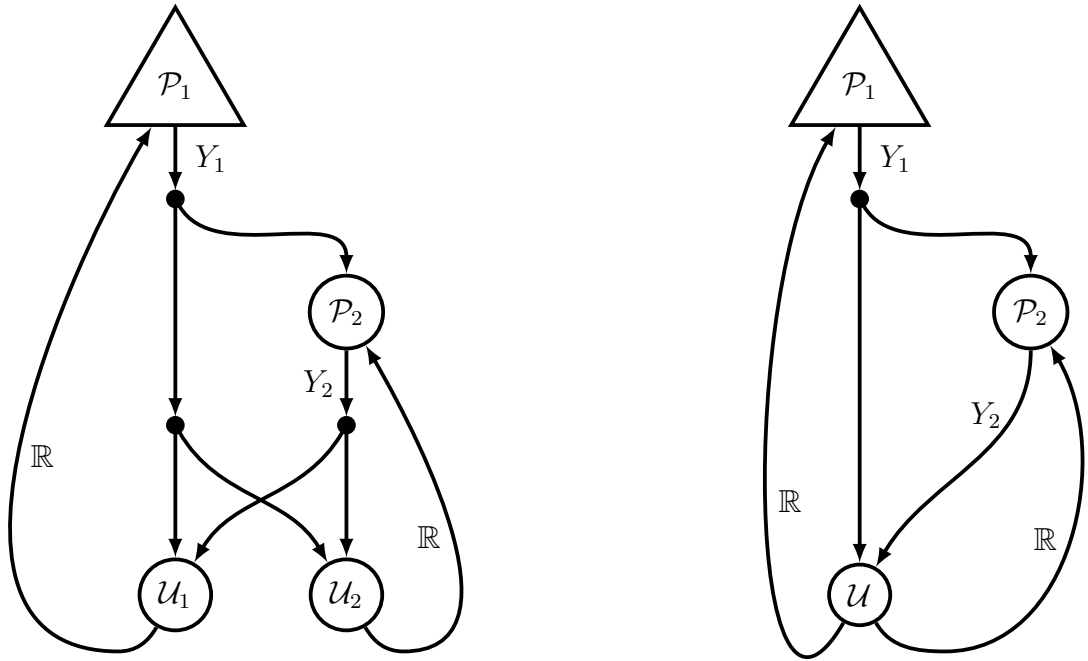

\stringdiagram{
	\node [triangle,draw] (P1) at (0,10){$\mathcal{P}_1$};
	\node [circle,scale = 0.5, fill=black,draw] (bP1) at (0,8) {};
	\node [circle,draw] (P2) at (3,6){$\mathcal{P}_2$};
	\node [circle,scale = 0.5, fill=black,draw] (uU1) at (0,4) {};
	\node [circle,scale = 0.5, fill=black,draw] (uU2) at (3,4) {};
	\node [circle,draw] (U1) at (0,1){$\mathcal U_1$};
	\node [circle,draw] (U2) at (3,1){$\mathcal U_2$};

	\draw [->] (P1) to node [right = 0.1cm, midway] {$Y_1$}   (bP1);
	\draw [->] (bP1) to [out = -60, in = 90] (P2);
	\draw [->] (bP1) to [out = -90, in = 90] (uU1);
	\draw [->] (uU1) to [out = -90, in = 90] (U1);
	\draw [->] (uU1) to [out = -60, in = 120] (U2);

	\draw [->] (P2) to [out=-90,in=90] node [left = 0cm, midway] {$Y_2$}   (uU2);
	\draw [->] (uU2) to [out = 240, in = 60] (U1);
	\draw [->] (uU2) to [out = -90, in = 90] (U2);
	\draw [->] (U1) to [out=-120, in=0] (-1,0) to [out=180,in=240] node [right] {$\R$} (P1);
	\draw [->] (U2) to [out=-60, in=180] (4,0) to [out=0,in=-60] node [left] {$\R$} (P2);

	\node [triangle,draw] (P1d) at (12,10){$\mathcal{P}_1$};
	\node [circle,scale = 0.5, fill=black,draw] (bP1d) at (12,8) {};
	\node [circle,draw] (P2d) at (15,6){$\mathcal{P}_2$};
	\node [circle,draw] (U1d) at (12,1){$\mathcal U$};

	\draw [->] (P1d) to node [right = 0.1cm, midway] {$Y_1$}   (bP1d);
	\draw [->] (bP1d) to [out = -90, in = 90] (U1d);
	\draw [->] (bP1d) to [out = -60, in = 90] (P2d);
	\draw [->] (P2d) to [out=-90,in=60] node [left = 0cm, near start] {$Y_2$}   (U1d);
	\draw [->] (U1d) to [out=-120, in=-120] node [right] {$\R$} (P1d);
	\draw [->] (U1d) to [out=-60, in=180] (13.5,0) to [out=0,in=-60] node [left] {$\R$} (P2d);
}
\caption{Ultimatum game}
\label{fig:seq_bas}
\end{figure}

The general form of such a game is depicted on the left of figure \ref{fig:seq_bas}.
As the structure of the string diagram shows, player 1 moves first and then the choice of player 1 is perfectly observed by player 2.
Observe that below the players, the diagram is the same as the general bimatrix game in figure \ref{fig:batt}: this is the general form of utility functions.

Alternatively the utility functions can be combined into a single utility function $\mathcal U : Y_1 \times Y_2 \to \R^2$ where
\[ \mathcal U (y_1, y_2) = \begin{cases} (y_1, N - y_1) & \text{ if } y_2 = A \\ (0, 0) & \text{ if } y_2 = B \end{cases} \]
This trivial reformulation results in a slightly different string diagram, depicted on the right of figure \ref{fig:seq_bas}.

These two string diagrams denote the same closed game $\G : I \to I$.
The strategy profiles of this game are
\[ \Sigma_\G = Y_1 \times (Y_1 \to Y_2) \]
so a strategy profile is a pair $(\sigma_1, \sigma_2)$, where $\sigma_1$ is a choice for player 1, and $\sigma_2$ is a choice for player 2 contingent on $Y_1$.
The equilibria defined by the relation $\E$ are precisely the pure strategy subgame-perfect equilibria, that is, those $(\sigma_1, \sigma_2)$ satisfying:
\begin{align*}
\mathcal U_1 (\sigma_1, \sigma_2 (\sigma_1)) &= \max_{y_1 \in Y_1} \mathcal U_1 (y_1, \sigma_2 (y_1)) \\
\mathcal U_2 (y_1, \sigma_2 (y_1)) &= \max_{y_2 \in Y_2} \mathcal U_2 (y_1, y_2) \text{ for all } y_1 \in Y_1
\end{align*}
In the example of the ultimatum game there are two equilibria, namely $(0, \sigma_2)$ and $(1, \sigma_2')$, where
\[ \sigma_2 (y_1) = R \text{ for all } y_1 \]
and
\[ \sigma_2' (y_1) = \begin{cases}
	R & \text{ if } y_1 = 0 \\
	A & \text{ if } y_1 > 0
\end{cases} \]

\subsection{Boxing}
The algebraic expressions that represent computations, decisions, or open games can be substituted by placeholders. Graphically, a substitution corresponds to boxing elements of a given open game (see Section \ref{sec:substitution}).  

\begin{figure}[H]
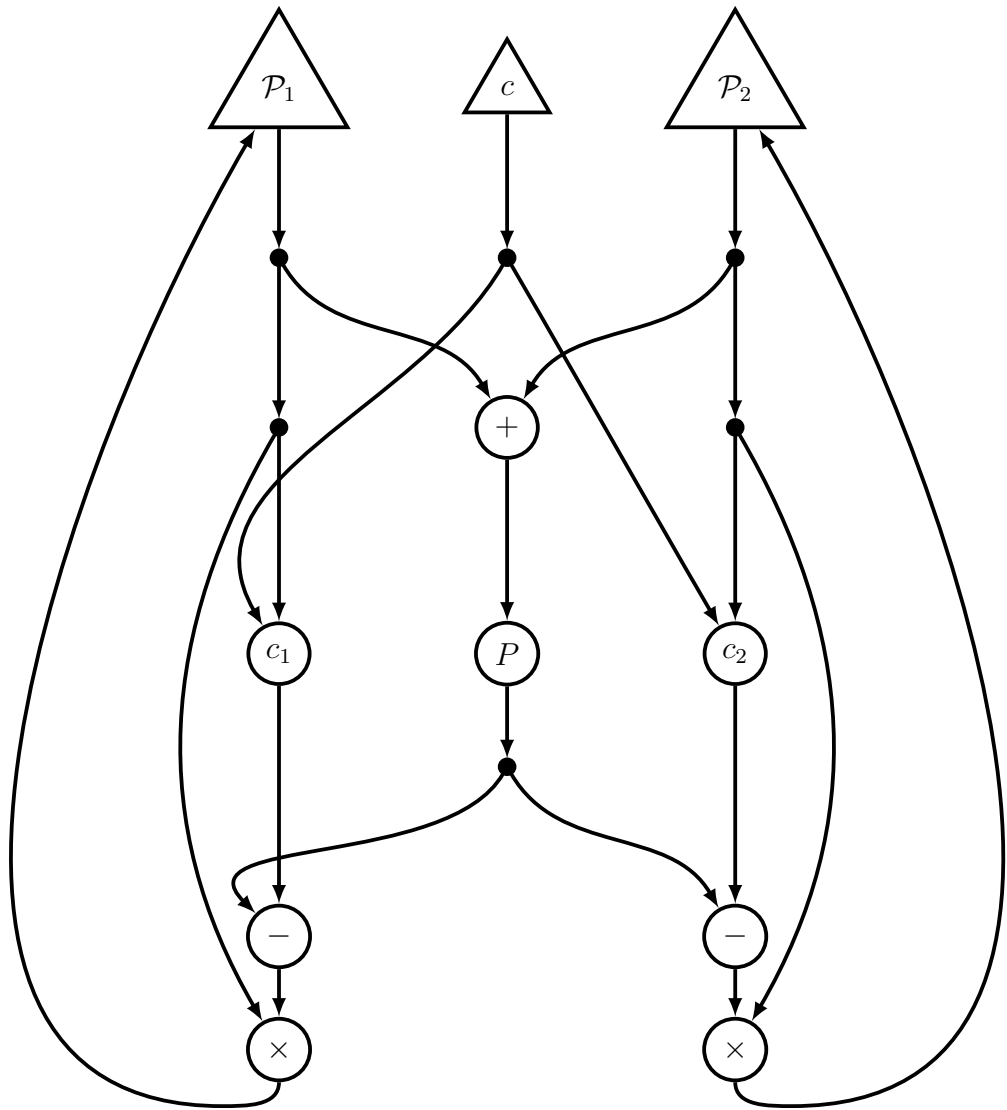

\begin{center}
\stringdiagram{
	\node [triangle, draw] (P1) at (0,18) {$\mathcal P_1$}; 
	\node [triangle, draw] (P2) at (8,18) {$\mathcal P_2$};
	\node [circle, scale=0.5, fill=black, draw] (m1) at (0,15) {}; 
	\node [circle, scale=0.5, fill=black, draw] (m2) at (8,15) {};
	\node [circle, scale=0.5, fill=black, draw] (m3) at (0,12) {}; 
	\node [circle, scale=0.5, fill=black, draw] (m4) at (8,12) {};
	\node [circle, draw] (a) at (4,12) {$+$}; 
	\node [circle, draw] (p) at (4,8) {$P$};
	\node [triangle, draw] (c) at (4,18) {$c$};
	\node [circle, scale=0.5, fill=black, draw] (cc) at (4,15) {};
	\node [circle, draw] (c1) at (0,8) {$c_1$}; 
	\node [circle, draw] (c2) at (8, 8) {$c_2$};
	\node [circle, scale=0.5, fill=black, draw] (m5) at (4,6) {};
	\node [circle, draw] (s1) at (0,3) {$-$}; 
	\node [circle, draw] (s2) at (8,3) {$-$};
	\node [circle, draw] (t1) at (0,1) {$\times$}; 
	\node [circle, draw] (t2) at (8,1) {$\times$};
	\draw [->] (P1) to [out=-90, in=90] (m1); 
	\draw [->] (P2) to [out=-90, in=90] (m2);
	\draw [->] (m1) to [out=-90, in=90] (m3); 
	\draw [->] (m1) to [out=-60, in=120] (a);
	\draw [->] (m2) to [out=240, in=60] (a); 
	\draw [->] (m2) to [out=-90, in=90] (m4);
	\draw [->] (m3) to [out=-90, in=90] (c1); 
	\draw [->] (m4) to [out=-90, in=90] (c2);
	\draw [->] (a) to [out=-90, in=90] (p);
	\draw [->] (p) to (m5);
	\draw [->] (c) to [out=-90, in=90] (cc);
	\draw [->] (cc) to [out=240, in=120] (c1);
	\draw [->] (cc) to [out=-60, in=120] (c2);
	\draw [->] (m5) to [out=240, in=135] (s1); 
	\draw [->] (c1) to [out=-90, in=90] (s1);
	\draw [->] (m5) to [out=-60, in=120] (s2); 
	\draw [->] (c2) to [out=-90, in=90] (s2);
	\draw [->] (m3) to [out=240, in=120] (t1); 
	\draw [->] (s1) to [out=-90, in=90] (t1);
	\draw [->] (s2) to [out=-90, in=90] (t2); 
	\draw [->] (m4) to [out=-60, in=60] (t2);
	\draw [->] (t1) to [out=-90, in=0] (-1, 0) to [out=180, in=240] (P1);
	\draw [->] (t2) to [out=-90, in=180] (9, 0) to [out=0, in=-60] (P2);
}
\caption{Cournot Duopoly}
\label{fig:cournot}
\end{center}
\end{figure}

The purpose of boxing is to hide information such that a focus on the crucial elements of the analysis is facilitated. For illustrating this point, we consider the Cournot duopoly \cite[p.389]{Mas-Colell1995}. 
The game is about two companies $\mathcal P_1$ and $ \mathcal P_2$ in a competitive market simultaneously deciding on the quantities $q_1$ and $q_2$ given an inverse demand function $P$ 
and some marginal costs c in the cost functions $c_1(c,q_1) $ and $c_2(c,q_2)$.
The profits $\pi_1$ and $\pi_2$ are given by
\begin{eqnarray}
\pi_1 &=& (P(q_1+q_2)-c_1(c,q_1))q_1\label{eq:doupoly_payoff1}\\
\pi_2 &=& (P(q_1+q_2)-c_2(c,q_2))q_2\label{eq:doupoly_payoff2}
\end{eqnarray}
Once we specify the model by $P=a-b(q_1+q_2)$, $c_1=c q_1$ such that $\pi_1=(a-b(q_1+q_2)-c)q_1$ (similarly $c_2$ and $\pi_2$)
the Nash equilibrium is given by 
\begin{equation}\label{eq:duopoly_nash}
q_1=q_2=(a-c)/3b
\end{equation}
The game, the companies, the profit, demand and cost functions are given in Figure \ref{fig:cournot}.  

Before giving the algebraic description and meaning of this string diagram, we will first simplify it.

While the exact details of the profit function may important for some questions, we focus on the timing of moves. For that reason we will hide the details by boxing the profit function $\pi$. 

\begin{figure}[H]
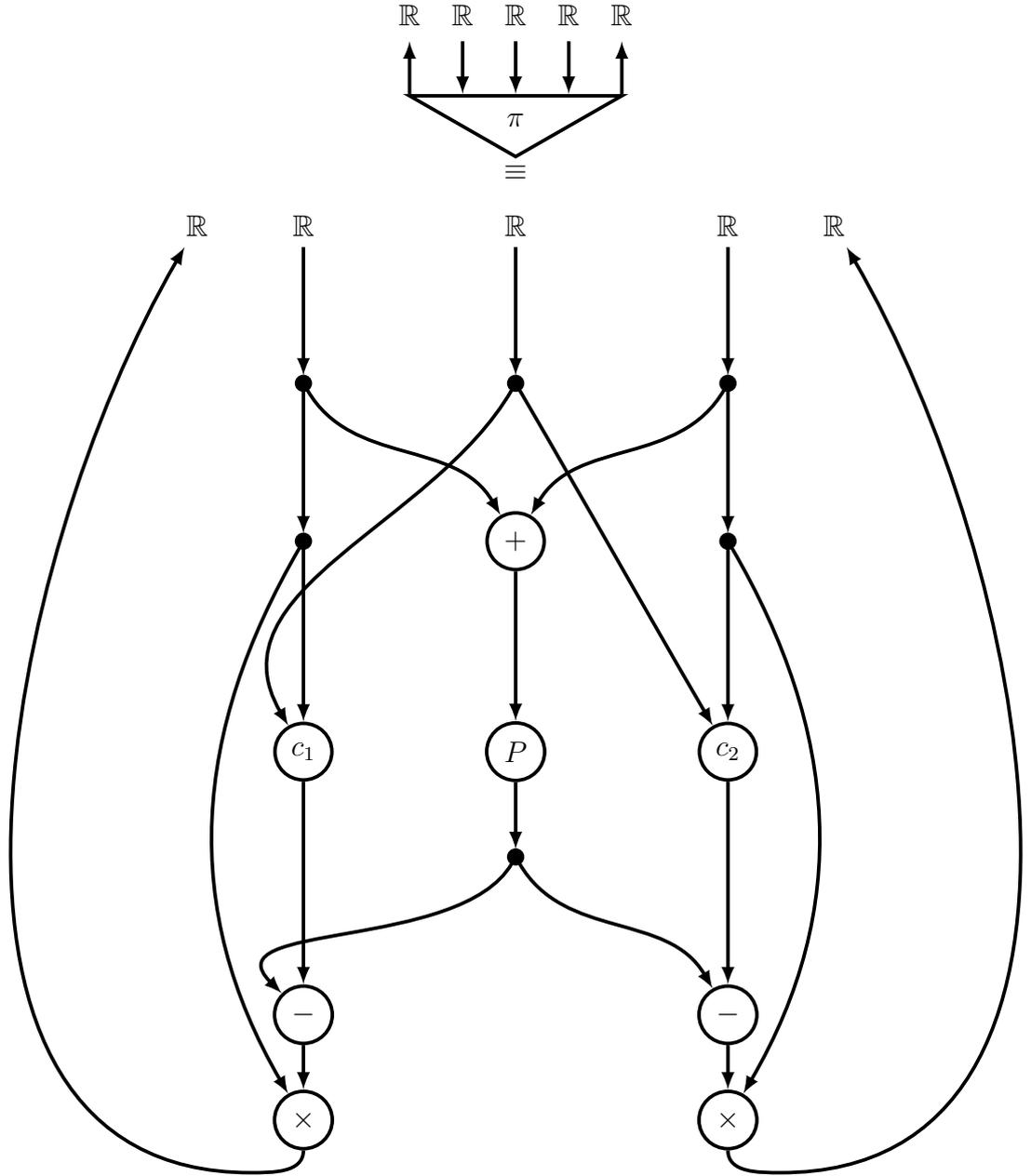

\stringdiagram{
	\node (R1e) at (2,22) {$\R$}; 
	\node (R2e) at (3,22) {$\R$}; 
	\node (R3e) at (4,22) {$\R$}; 
	\node (R4e) at (5,22) {$\R$}; 
	\node (R5e) at (6,22) {$\R$};
	\node [isosceles triangle, isosceles triangle apex angle = 120, minimum width = 3cm, shape border rotate=270, draw] (def) at (4, 20) {$\pi$};

	\draw [->] [transform canvas = {xshift= -1.5cm}] (4,20.5) to (4,21.5);
	\draw [->] [transform canvas = {xshift= -0.75cm}] (4,21.5) to (4,20.5);
	\draw [->] [transform canvas = {xshift= 0cm}] (4,21.5) to (4,20.5);
	\draw [->] [transform canvas = {xshift= 0.75cm}] (4,21.5) to (4,20.5);
	\draw [->] [transform canvas = {xshift= 1.5cm}] (4,20.5) to (4,21.5);

	\node (e) at (4, 19) {$\equiv$};
	
	\node (R1) at (0, 18) {$\R$}; \node (R2) at (-2,18) {$\R$}; \node (R3) at (4,18) {$\R$}; \node (R4) at (8,18) {$\R$}; \node (R5) at (10,18) {$\R$};

	\node [circle, scale=0.5, fill=black, draw] (m1) at (0,15) {}; 
	\node [circle, scale=0.5, fill=black, draw] (m2) at (8,15) {};
	\node [circle, scale=0.5, fill=black, draw] (m3) at (0,12) {}; 
	\node [circle, scale=0.5, fill=black, draw] (m4) at (8,12) {};
	\node [circle, draw] (a) at (4,12) {$+$}; 
	\node [circle, draw] (p) at (4,8) {$P$};
	\node [circle, scale=0.5, fill=black, draw] (cc) at (4,15) {};
	\node [circle, draw] (c1) at (0,8) {$c_1$}; 
	\node [circle, draw] (c2) at (8,8) {$c_2$};
	\node [circle, scale=0.5, fill=black, draw] (m5) at (4,6) {};
	\node [circle, draw] (s1) at (0,3) {$-$}; 
	\node [circle, draw] (s2) at (8,3) {$-$};
	\node [circle, draw] (t1) at (0,1) {$\times$}; 
	\node [circle, draw] (t2) at (8,1) {$\times$};
	\draw [->] (R1) to [out=-90, in=90] (m1); 
	\draw [->] (R4) to [out=-90, in=90] (m2);
	\draw [->] (m1) to [out=-90, in=90] (m3); 
	\draw [->] (m1) to [out=-60, in=120] (a);
	\draw [->] (m2) to [out=240, in=60] (a); 
	\draw [->] (m2) to [out=-90, in=90] (m4);
	\draw [->] (m3) to [out=-90, in=90] (c1); 
	\draw [->] (m4) to [out=-90, in=90] (c2);
	\draw [->] (a) to [out=-90, in=90] (p); 
	\draw [->] (p) to (m5);
	\draw [->] (R3) to [out=-90, in=90] (cc);
	\draw [->] (cc) to [out=240, in=120] (c1);
	\draw [->] (cc) to [out=-60, in=120] (c2);

	\draw [->] (m5) to [out=240, in=135] (s1); 
	\draw [->] (c1) to [out=-90, in=90] (s1);
	\draw [->] (m5) to [out=-60, in=120] (s2); 
	\draw [->] (c2) to [out=-90, in=90] (s2);
	\draw [->] (m3) to [out=240, in=120] (t1); 
	\draw [->] (s1) to [out=-90, in=90] (t1);
	\draw [->] (s2) to [out=-90, in=90] (t2); 
	\draw [->] (m4) to [out=-60, in=60] (t2);
	\draw [->] (t1) to [out=-90, in=0] (-1, 0) to [out=180, in=240] (R2);
	\draw [->] (t2) to [out=-90, in=180] (9, 0) to [out=0, in=-60] (R5);

}
\caption{Box of the Demand and Profit Functions in a Duopoly Game}
\label{fig:cournot_box}
\end{figure}

The algebraic definition of $\pi$ has a diagrammatic representation in \ref{fig:cournot_box}.

In Figure \ref{fig:cournot_box} we have defined an open game $\pi : \R^* \otimes \R^{\otimes 3} \otimes \R^*$.
The equivalent algebraic expression, which (with some practice) can be read from the string diagram, is
\[ \pi = \tau_\R^{\otimes 2} \circ (\sigma_{\R^*, \R} \otimes \R \otimes \R^*) \circ (\R^* \otimes f \otimes \R^*) \]
where
\begin{align*}
	f &: \R^{\otimes 3} \to \R^{\otimes 6} \\
	f &= \times^{\otimes 2} \circ (\R \otimes g \otimes \R) \circ (\Delta_\R \otimes \R \otimes + \otimes \Delta_\R) \circ (\R \otimes \sigma_{\R, \R}^{\otimes 2} \otimes R) \circ \Delta_\R^{\otimes 3} \\
	g &: \R^{\otimes 5} \to \R^{\otimes 5} \\
	g &= -^{\otimes 2} \circ (\sigma_{\R, \R} \otimes \R^{\otimes 2}) \circ (\R \otimes \Delta_\R \otimes \R) \circ (c_1 \otimes P \otimes c_2) \circ (\sigma_{\R, \R} \otimes \R^{\otimes 3})
\end{align*}
The meaning of $\pi$ is the 4-tuple $(\Sigma_\pi, \P_\pi, \C_\pi, \E_\pi)$ where
\begin{itemize}
	\item $\Sigma_\pi = I$
	\item $\P_\pi (\bullet, (q_1, c, q2)) = \bullet$
	\item $\C_\pi(\bullet,(q_1,c,q_2),\bullet)=( (P(q_1+q_2) -c_1(c,q_1))q_1 , (P(q_1+q_2) -c_2(c,q_2))q_2)$
	\item $\E_\pi ((q_1, c, q_2), k) = \{ \bullet \}$
\end{itemize}
Thus the only non trivial part is the coplay function which is precisely the payoff functions in equations (\ref{eq:doupoly_payoff1}), (\ref{eq:doupoly_payoff2}).

\begin{figure}[H]
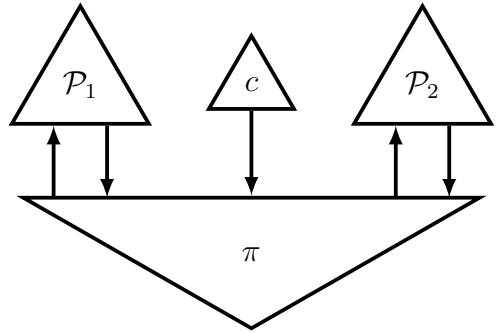

\stringdiagram{
	\node [triangle, draw] (P1) at (0, 3) {$\mathcal P_1$};  \node [triangle, draw] (P2) at (6, 3) {$\mathcal P_2$};
	\node [triangle, draw] (b) at (3,3) {$c$};
	\node [isosceles triangle, isosceles triangle apex angle = 120, minimum width = 6cm, shape border rotate=270, draw] (pi) at (3, 0) {$\pi$};
	\draw [->] [transform canvas = {xshift=10pt}] (P1) to (0,1);
	\draw [->] [transform canvas = {xshift=-10pt}] (0,1) to (P1);

	\draw [->] [transform canvas = {xshift=10pt}] (P2) to (6,1);
	\draw [->] [transform canvas = {xshift=-10pt}] (6,1) to (P2);

	\draw [->] (b) to [out=-90, in=90] (pi);
}
\caption{Cournot Duopoly}
\label{fig:cournot_boxed}
\end{figure}

Taking these functions as being hidden in a box simplifies the string diagram \ref{fig:cournot} into \ref{fig:cournot_boxed}.

Let $\mathcal G = \pi \circ (\mathcal P_1 \otimes c \otimes \mathcal P_2) : I \to I$ be the game denoted in Figure \ref{fig:cournot_boxed}. This game has trivial play and coplay functions, but its set of strategy profiles is $\Sigma_{\mathcal G} = \R^2$, and the function $\mathcal E$ gives the set of Nash equilibria. For example, assuming the same functional form, we have
\[ \mathbb E_{\mathcal G} (\bullet, \bullet) = \left\{ \left( \frac{a-c}{3b}, \frac{a-c}{3b} \right) \right\} \]

\subsection{Variants of Timing}
\begin{figure}[H]
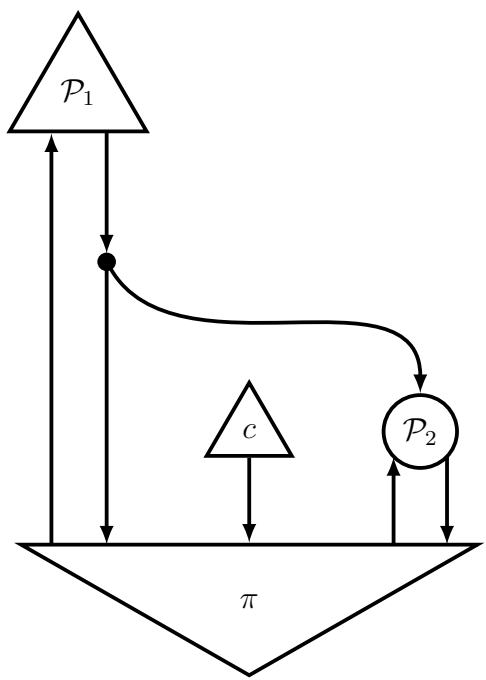

\stringdiagram{
	\node [triangle, draw] (P1) at (0, 9) {$\mathcal P_1$};  \node [circle,draw] (P2) at (6, 3) {$\mathcal P_2$};
	\node [triangle, draw] (b) at (3,3) {$c$};
	\node [circle, scale=0.5, fill=black, draw] (m1) at (0.5,6) {}; 
	\node [isosceles triangle, isosceles triangle apex angle = 120, minimum width = 6cm, shape border rotate=270, draw] (pi) at (3, 0) {$\pi$};
	\draw [->] [transform canvas = {xshift=0pt}] (0.5,8.3) to (m1);
	\draw [->] [transform canvas = {xshift=0pt}] (m1) to (0.5,1);
	\draw [->] [transform canvas = {xshift=-10pt}] (0,1) to (P1);

	\draw [->] [transform canvas = {xshift=10pt}] (6,2.55) to (6,1);
	\draw [->] [transform canvas = {xshift=-10pt}] (6,1) to (6,2.55);

	\draw [->] (b) to [out=-90, in=90] (pi);
	\draw [->] (m1) to [out=-60, in=90] (P2);
}
\caption{Stackelberg Duopoly}
\label{fig:stackelberg}
\end{figure}

In the analysis of a duopoly, the timing is crucial. So, we may be interested in comparing the Cournot model to a model where there is a first-mover, a Stackelberg duopoly, \cite[p.67]{Fudenberg1991}. As the profit function does not change, we can reuse this box. The string diagram is represented in  Figure \ref{fig:stackelberg}.

Comparing this Figure to the simplified Cournot diagram before, makes clear what the focus of the analysis is: the timing of the decision. 

\begin{figure}[H]
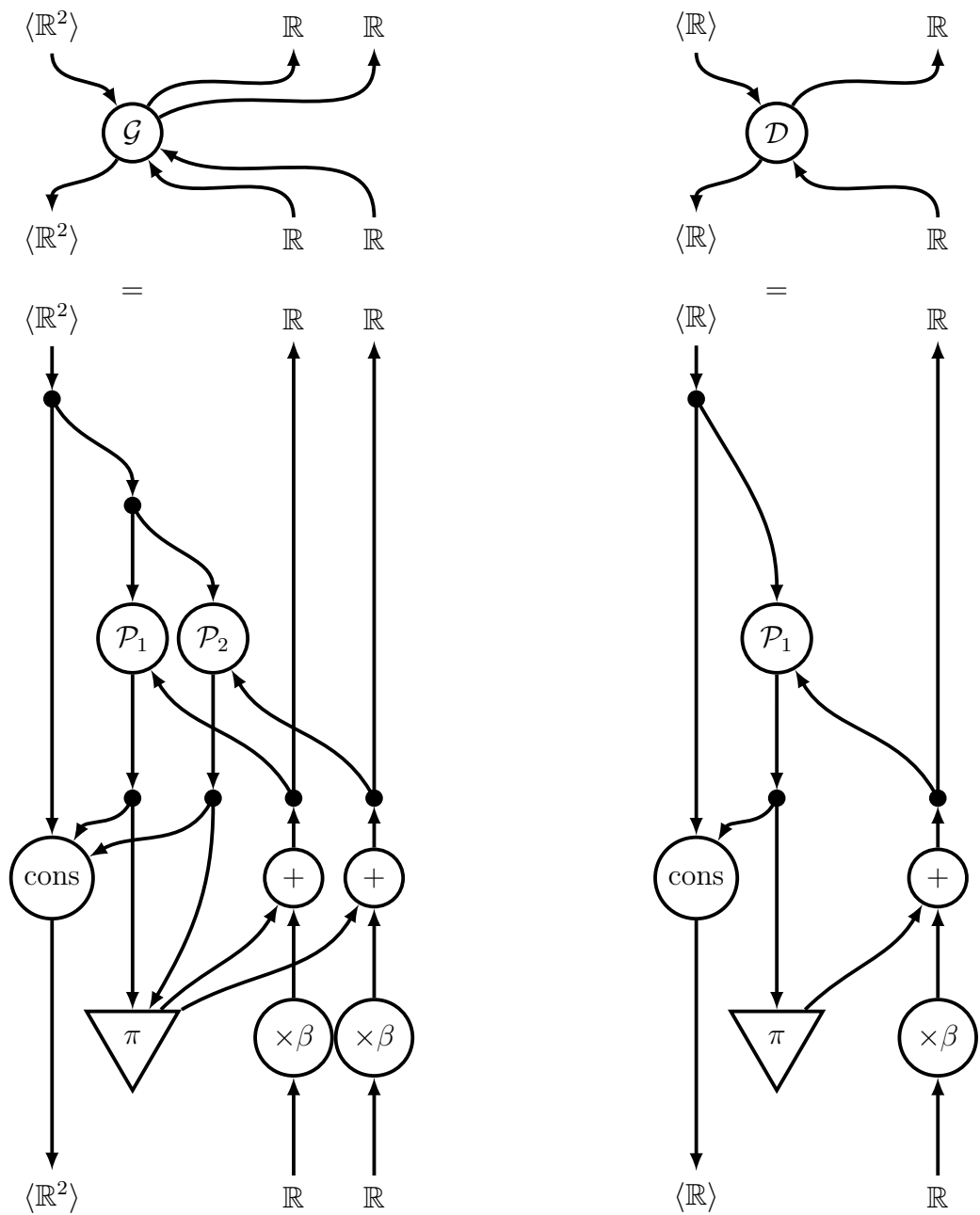

\stringdiagram{
	\node (Xe) at (0,22) {$\left< \R^2\right>$};
	\node (R1e) at (4.5,22) {$\R$};
	\node (R2e) at (6,22) {$\R$};
	\node (Ye) at (0,18) {$\left< \R^2\right>$};
	\node (S1e) at (4.5,18) {$\R$};
	\node (S2e) at (6,18) {$\R$};

	\node [circle, draw] (G) at (1.5,20) {$\mathcal{G}$};

	\draw [->] (Xe) to [out=-90, in=120] (G);
	\draw [->] (G) to [out=60, in=-90] (R1e);
	\draw [->] (G) to [out=30, in=-90] (R2e);
	\draw [->] (G) to [out=240, in=90] (Ye);
	\draw [->] (S1e) to [out=90, in=-60] (G);
	\draw [->] (S2e) to [out=90, in=-30] (G);

	\node (e1) at (1.5,17) {$=$};

	\node (X) at (0,16.5) {$\left< \R^2\right>$};
	\node (R1) at (4.5,16.5) {$\R$};
	\node (R2) at (6,16.5) {$\R$};
	\node (Y) at (0,0) {$\left< \R^2\right>$};
	\node (S1) at (4.5,0) {$\R$};
	\node (S2) at (6,0) {$\R$};

	\node [inverted triangle, draw] (U) at (1.5,3) {$\pi$};
	\node [circle, draw] (b1) at (4.5,3) {$\times \beta$};
	\node [circle, draw] (b2) at (6,3) {$\times \beta$};
	\node [circle, draw] (p1) at (4.5,6) {$+$};
	\node [circle, draw] (p2) at (6,6) {$+$};
	\node [circle, scale=0.5, fill=black, draw] (c5) at (4.5,7.5) {};
	\node [circle, scale=0.5, fill=black, draw] (c6) at (6,7.5) {};
	\node [circle, draw] (cons) at (0,6) {cons};
	\node [circle, scale=0.5, fill=black, draw] (c3) at (1.5,7.5) {};
	\node [circle, scale=0.5, fill=black, draw] (c4) at (3,7.5) {};
	\node [circle, draw] (P1) at (1.5,10.5) {$\mathcal{P}_1$};
	\node [circle, draw] (P2) at (3,10.5) {$\mathcal{P}_2$};
	\node [circle, scale=0.5, fill=black, draw] (c2) at (1.5,13.) {};
	\node [circle, scale=0.5, fill=black, draw] (c1) at (0,15) {};

	\draw [->] (X) to [out=-90, in=90] (c1);
	\draw [->] (c1) to [out=-90, in=90] (cons);
	\draw [->] (c1) to [out=-60, in=90] (c2);
	\draw [->] (c2) to [out=-90, in=90] (P1);
	\draw [->] (c2) to [out=-60, in=90] (P2);
	\draw [->] (P1) to [out=-90, in=90] (c3);
	\draw [->] (P2) to [out=-90, in=90] (c4);
	\draw [->] (c3) to [out=-120, in=60] (cons);
	\draw [->] (c3) to [out=-90, in=90] (U);
	\draw [->] (c4) to [out=-120, in=30] (cons);
	\draw [->] (c4) to [out=-90, in=60] (U);
	\draw [->] (cons) to [out=-90, in=90] (Y);
	\draw [->] (S1) to [out=90, in=-90] (b1);
	\draw [->] (S2) to [out=90, in=-90] (b2);
	\draw [->] (b1) to [out=90, in=-90] (p1);
	\draw [->] (b2) to [out=90, in=-90] (p2);
	\draw [->] (p1) to [out=90, in=-90] (c5);
	\draw [->] (p2) to [out=90, in=-90] (c6);
	\draw [->] (c5) to [out=90, in=-90] (R1);
	\draw [->] (c6) to [out=90, in=-90] (R2);
	\draw [->] (c5) to [out=120, in=-60] (P1);
	\draw [->] (c6) to [out=120, in=-60] (P2);
	\draw [->] (U) to [out=45, in=-120] (p1);
	\draw [->] (U) to [out=30, in=-120] (p2);
	
	\node (Xde) at (12,22) {$\left< \R\right>$};
	\node (R1de) at (16.5,22) {$\R$};
	\node (Yde) at (12,18) {$\left< \R\right>$};
	\node (S1de) at (16.5,18) {$\R$};
	\node [circle, draw] (D) at (13.5,20) {$\mathcal{D}$};

	\draw [->] (Xde) to [out=-90, in=120] (D);
	\draw [->] (D) to [out=60, in=-90] (R1de);
	\draw [->] (D) to [out=240, in=90] (Yde);
	\draw [->] (S1de) to [out=90, in=-60] (D);

	\node (e2) at (13.5,17) {$=$};

	\node (Xd) at (12,16.5) {$\left< \R\right>$};
	\node (R1d) at (16.5,16.5) {$\R$};
	\node (Yd) at (12,0) {$\left< \R\right>$};
	\node (S1d) at (16.5,0) {$\R$};

	\node [inverted triangle, draw] (Ud) at (13.5,3) {$\pi$};
	\node [circle, draw] (b1d) at (16.5,3) {$\times \beta$};
	\node [circle, draw] (p1d) at (16.5,6) {$+$};
	\node [circle, scale=0.5, fill=black, draw] (c5d) at (16.5,7.5) {};
	\node [circle, draw] (consd) at (12,6) {cons};
	\node [circle, scale=0.5, fill=black, draw] (c3d) at (13.5,7.5) {};
	\node [circle, draw] (P1d) at (13.5,10.5) {$\mathcal{P}_1$};
	\node [circle, scale=0.5, fill=black, draw] (c1d) at (12,15) {};

	\draw [->] (Xd) to [out=-90, in=90] (c1d);
	\draw [->] (c1d) to [out=-90, in=90] (consd);
	\draw [->] (c1d) to [out=-60, in=90] (P1d);
	\draw [->] (P1d) to [out=-90, in=90] (c3d);
	\draw [->] (c3d) to [out=-120, in=60] (consd);
	\draw [->] (c3d) to [out=-90, in=90] (Ud);
	\draw [->] (consd) to [out=-90, in=90] (Yd);
	\draw [->] (S1d) to [out=90, in=-90] (b1d);
	\draw [->] (b1d) to [out=90, in=-90] (p1d);
	\draw [->] (p1d) to [out=90, in=-90] (c5d);
	\draw [->] (c5d) to [out=90, in=-90] (R1d);
	\draw [->] (c5d) to [out=120, in=-60] (P1d);
	\draw [->] (Ud) to [out=45, in=-120] (p1d);

}
\caption{Box of the Repeated Cournot Duopoly and Decision}
\label{fig:repeat_box}
\end{figure}

\begin{figure}[H]
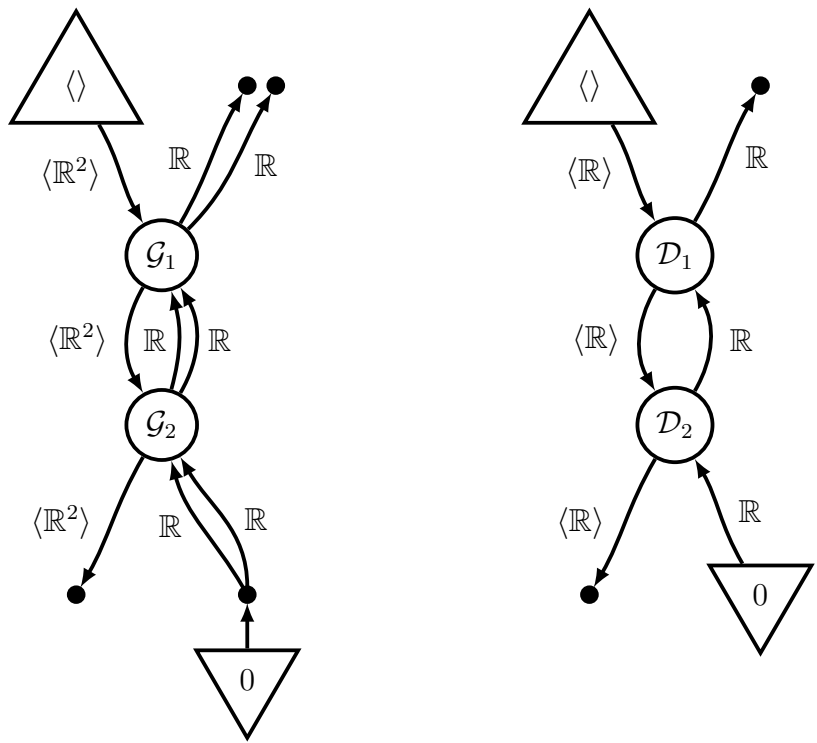

\stringdiagram{
	\node [triangle, draw] (X) at (0,9) {$\left< \right>$};
	\node [circle, scale=0.5, fill=black, draw]  (R1) at (3,9) {}; 
	\node [circle, scale=0.5, fill=black, draw]  (R2) at (3.5,9) {}; 
	\node [circle, scale=0.5, fill=black, draw] (Y) at (0,0) {};
	\node [circle, scale=0.5, fill=black, draw] (c1) at (3,0) {};
	\node [inverted triangle, draw] (U) at (3,-1.5) {0};
	\node [circle, draw] (G1) at (1.5,6) {$\mathcal{G}_1$};
	\node [circle, draw] (G2) at (1.5,3) {$\mathcal{G}_2$};

	\draw [->] (X) to [out=-60, in=120] node [left = 0.1cm, midway] {$\left< \R^2\right>$}  (G1);
	\draw [->] (G1) to [out=-120, in=120] node [left = 0.1cm, midway] {$\left<\R^2\right>$}  (G2);
	\draw [->] (G2) to [out=-120, in=60] node [left = 0.1cm, midway] {$\left<\R^2\right>$}  (Y);
	\draw [->] (U) to [out=90, in=-90]   (c1);
	\draw [->] (c1) to [out=120, in=-75] node [left = 0.1cm, midway] {$\R$}  (G2);
	\draw [->] (c1) to [out=90, in=-60] node [right = 0.1cm, midway] {$\R$}  (G2);
	\draw [->] (G2) to [out=75, in=-75] node [left = 0cm, midway] {$\R$}  (G1);
	\draw [->] (G2) to [out=60, in=-60] node [right = 0cm, midway] {$\R$}  (G1);
	\draw [->] (G1) to [out=60, in=-120]node [left = 0.1cm, midway] {$\R$}   (R1);
	\draw [->] (G1) to [out=45, in=-120]node [right = 0.1cm, midway] {$\R$}   (R2);

	\node [triangle, draw] (Xd) at (9,9) {$\left< \right>$};
	\node [circle, scale=0.5, fill=black, draw] (Rd) at (12,9) {};
	\node [circle, scale=0.5, fill=black, draw] (Yd) at (9,0) {};
	\node [inverted triangle, draw] (Ud) at (12,0) {0};
	\node [circle, draw] (G1d) at (10.5,6) {$\mathcal{D}_1$};
	\node [circle, draw] (G2d) at (10.5,3) {$\mathcal{D}_2$};

	\draw [->] (Xd) to [out=-60, in=120] node [left = 0.1cm, midway] {$\left< \R \right>$}  (G1d);
	\draw [->] (G1d) to [out=-120, in=120] node [left = 0.1cm, midway] {$\left< \R \right>$}  (G2d);
	\draw [->] (G2d) to [out=-120, in=60] node [left = 0.1cm, midway] {$\left< \R \right>$}  (Yd);
	\draw [->] (Ud) to [out=120, in=-60] node [right = 0.1cm, midway] {$\R$}  (G2d);
	\draw [->] (G2d) to [out=60, in=-60] node [right = 0.1cm, midway] {$\R$}  (G1d);
	\draw [->] (G1d) to [out=60, in=-120]node [right = 0.1cm, midway] {$\R$}   (Rd);
}
\caption{Repeated Cournot Duopoly and Decision}
\label{fig:repeat}
\end{figure}

Suppose, now we would like to consider potential dynamic effects. So, what would happen, if we considered a two-time repeated Cournot duopoly?
Figure \ref{fig:repeat_box}  represents the stage game in detail (bottom left) and boxed (top left) where $\left< \R^2\right>$ denotes a list of pairs of reals.
Figure \ref{fig:repeat} represents the repeated game by connecting two boxed stage games. In this Figure as well as in Figure \ref{fig:repeat_box} we have also included a repeated decision problem, stage representation as well as the repeated game using the boxed stage game. What should become obvious by direct comparisons is the parallelnes of constructions that are used for a decision for a two-player game, or (not shown here) for an n-player game.   

\subsection{Exogenous and Endogeous Variables}
So far we have limited the composition of players to be either only parallel or sequential. Next, we build up a game using both operations.

\begin{figure}[H]
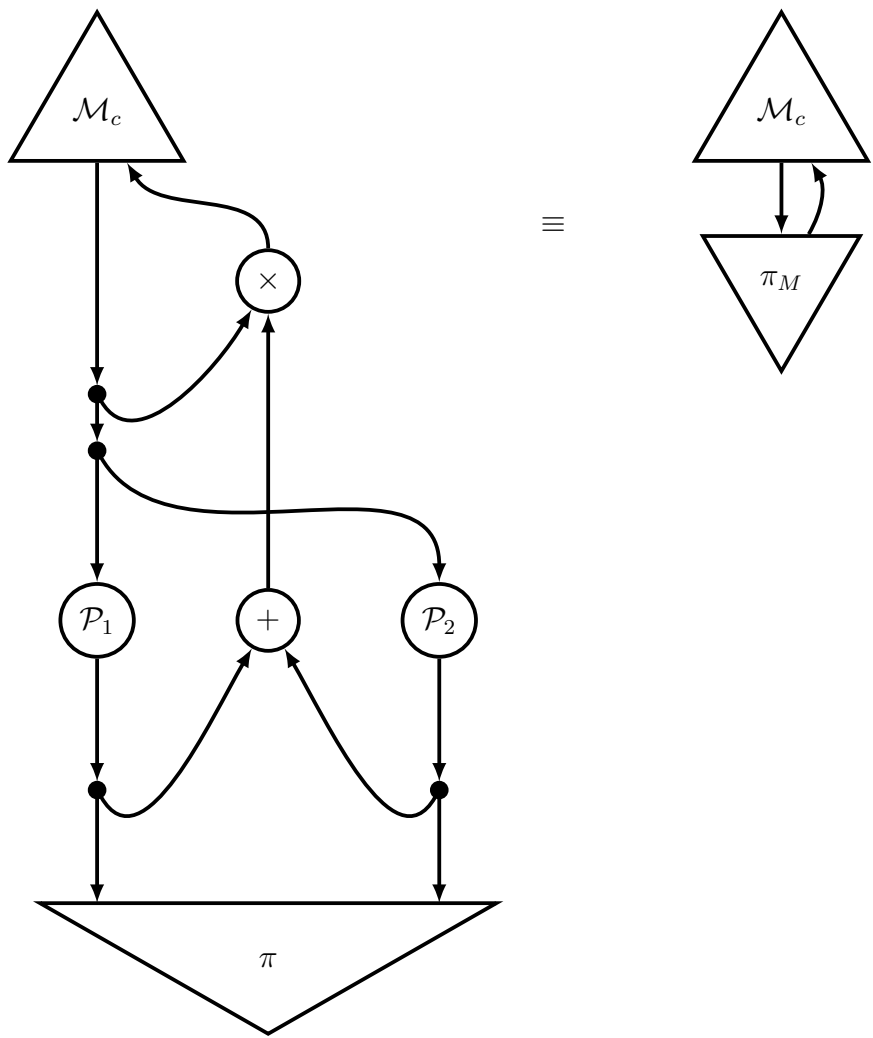

\stringdiagram{
	\node [triangle, draw] (Mc) at (0,15) {$\mathcal{M}_c$};
	\node [circle, scale=0.5, fill=black, draw] (m1) at (0,10) {};
	\node [circle, scale=0.5, fill=black, draw] (m2) at (0,9) {};
	\node [circle, draw] (P1) at (0,6) {$\mathcal P_1$}; 
	\node [circle, draw] (P2) at (6,6) {$\mathcal P_2$};
	\node [circle, scale=0.5, fill=black, draw] (m4) at (0,3) {};
	\node [circle, draw] (p) at (3,6) {$+$};
	\node [circle, scale=0.5, fill=black, draw] (m5) at (6,3) {};
	\node [circle, draw] (t) at (3,12) {$\times$};
	\node [isosceles triangle, isosceles triangle apex angle = 120, minimum width = 6cm, shape border rotate=270, draw] (pi) at (3, 0) {$\pi$};
	\draw [->] (t) to [out=90, in=-60] (Mc);
	\draw [->] (Mc) to [out=-90, in=90] (m1);
	\draw [->] (m1) to [out=-60, in=-120] (t);
	\draw [->] (m1) to [out=-90, in=90] (m2);
	\draw [->] (m2) to [out=-90, in=90] (P1);
	\draw [->] (m2) to [out=-60, in=90] (P2); 
	\draw [->] (p) to [out=90, in=-90] (t);
	\draw [->] (m4) to [out=-60, in=-120] (p);
	\draw [->] (m5) to [out=240, in=-60] (p);
	\draw [->] (P1) to [out=-90, in=90] (m4);
	\draw [->] (P2) to [out=-90, in=90] (m5);
	\draw [->] [transform canvas = {xshift=0cm}] (0,3) to (0,1);
	\draw [->] [transform canvas = {xshift=4.5cm}] (0,3) to (0,1);
	\node at (8,13) {$\equiv$};
	\node [triangle, draw] (P3) at (12,15){$\mathcal{M}_c$};
	\node [inverted triangle, draw] (q3) at (12,12){$\pi_M$};
	\draw [->] (P3) to [out = -90, in = 90] (q3);
	\draw [->] (q3) to [out = 60, in = -60] (P3);
}
\caption{Upstream Monopolist and Downstream Duopoly}
\label{fig:taxing}
\end{figure}

In Figure \ref{fig:taxing} we see the corresponding string diagram of three players, 
a monopolist $\mathcal{M}_c$ moving first by setting the price of a good that is the input to the sequential second stage of the game where there is again a duopoly of two competing companies. 
This example shows how to endogeneize a parameter: in the duopoly game in Figure \ref{fig:cournot} we have set the marginal costs of the two companies to be exogenous.
Now we model this parameter to be the price of the input good of the monopolist $c=p_m$.
The two companies now have profit functions defined by $\pi_i=(a-b(q_1+q_2)) q_i - p_m q_i$ for $i=1, 2$.
The profit function of the monopolist is now $\pi=p_m(q_1+q_2)$.
The two companies move simultaneously by setting quantities as in the Cournot example before given by $q_1=q_2=(a-p_m)/(3b)$ while the monopolist sets the price at $p_m=a/2$.

This example highlights another dimension of the compositionality we get by string diagrams. 
In the duopoly examples above we have taken the parameter $c$ of the cost functions to be exogenous. 
Diagrammatically the triangle in Figure \ref{fig:cournot} represented this exogenous marginal cost parameter which is there a computation without inputs and feeds into the cost function $c_1$ and $c_2$.
Now this cost parameter is the monopoly price and the action of the monopolist. 

Note that the boxed $\pi_M$ is actually a game which contains strategic players even so from the perspective of the monopolist $\mathcal M_c$ it behaves just as another outcome function.

\section{Conclusion}
We have shown how to use string diagrams to represent and construct games in a compositional way. 
This graphical language complements the traditional extensive form representation which emphasizes the unfolding of a game when it is played.
As opposed to that the string diagrams highlight the network like structure of the parallel and sequential composition of the various decisions involved
and also allow for a hierarchical composition by boxing layers of games.
The examples we provide are simple but the mathematics on which string diagrams are based is complex.
So what do we actually gain by introducing string diagrams?

There are two aspects. 
First, it is important to emphasize, that most of the mathematical effort flows into establishing that the diagrammatic language is an exact representation of an underlying categorical treatment. 
In practice, modelling takes place at the level of the graphical language such that the underlying details can be safely ignored - 
exactly because the difficult work of establishing a formal semantics has been done once and for all.
The choice of a modelling language is an important one because it makes life simpler or harder for people who want to model. 
Graphical languages are typically easier to use than the formulas they represent. 
In physics and computation, for instance, string diagram have been a success story because they allow to abstract away complications and focus on the 
essentials.\footnote{Note that there are similar visual languages in the literature on multi-agent systems.
Bayesian networks use graphs for the representation of conditional independence assumption in joint densities of random variables.
Influence diagrams add decision and utility nodes to the Bayesian network for the representation of games.}

There is a general lesson to be learned - one that is well known in computer science but not so much in economics. 
Even if it is possible to model an object in a particular formal language, it may still be insightful for our understanding to represent the problem in a different language. 
Computer programmers usually do not want to be bothered with the procedures at the level of machine language when writing code. 
They want to work at a language level that makes understanding the used concepts simpler. 
The same lesson applies to game theory. 
The diagrammatic language refers to the same underlying objects as does standard game theory. 
But our claim is that having another layer of language available, is of help to our understanding and our modelling efforts.

There is a second aspect. 
In economic applications there is a growing need for computer assisted modelling of large systems. 
A recent example is the energy grid that is changing in some countries due to green energy policies. 
This poses a challenge for existing theory. 
On the one hand the theory cannot easily be extended by hand to accommodate such large systems
where the notion of a theory and the real implementation is blurred.
On the other hand an implementation of games in computer code starting at the usual mathematical language is often cumbersome and not scalable.

Modelling large system with the help of computers is a classical topic in computer science. 
Our framework is inspired and shares roots with the parts of computer science theory that are exactly concerned with these questions.
Hence, several hard learned lessons from computer science are reflected in our framework:

\begin{enumerate}
\item We have emphasized the notion of compositionality. 
This basically means that larger systems can be composed from subparts. 
It is one way of dealing with complexity of large software systems or in our case with potentially large games. 
\item Coping with complexity can be approached by the usage of different layers of languages and formal mathematical translations in between them.
\item We establish a close connection between the formal theory and the computer code: The string diagrams are directly translatable into the functional programming language Haskell. 
\item By translating into this setting a large set of powerful tools of software design and verification can be applied: automated equilibrium checking, predicate testing, modal logics, type and category theory etc.
\end{enumerate}


\bibliographystyle{plain}
\bibliography{../references}

\appendix
\section{The algebra of open games}

In this appendix we will explicitly give the 4-tuple $(\Sigma_\G, \P_\G, \C_\G, \E_\G)$ for the atomic games in section \ref{sec:Open_game_units} and the ways they are composed in section \ref{sec:composition_operations}.
This is enough to reconstruct the data for every example in this paper.
We will give minimal formal definitions; the details can be found in [jules' thesis] and \cite{Ghani_Hedges2016}.

Recall that if $\G : X \otimes S^* \to Y \otimes R^*$ then:
\begin{itemize}
	\item $\Sigma_\G$ is a set
	\item $\P_\G : \Sigma_\G \times X \to Y$
	\item $\C_\G : \Sigma_\G \times X \times R \to S$
	\item $\E_\G : X \times (Y \to R) \to \mathcal P (\Sigma_\G)$
\end{itemize}

Given this data for $\G : X \otimes T^* \to Y \otimes S^*$ and $\H : Y \otimes S^* \to Z \otimes R^*$, the composition
\[ \H \circ \G : X \otimes T^* \to Z \otimes R^* \]
is defined by:
\begin{itemize}
\item $\Sigma_{\H \circ \G} = \Sigma_\G \times \Sigma_\H$
\item $\P_{\H \circ \G} ((\sigma_1, \sigma_2), x) = \P_\H (\sigma_2, \P_\G (\sigma_1, x))$
\item $\C_{\H \circ \G} ((\sigma_1, \sigma_2), x, r) = \C_\G (\sigma_1, x, \C_\H (\sigma_2, \P_\G (\sigma_1, x), r))$
\item $\E_{\G \circ \H} (x, k) = \left\{ (\sigma_1, \sigma_2) \in \Sigma_\G \times \Sigma_\H \ \left|\ \begin{matrix}
	\sigma_1 \in \E_\G (x, k') \text{, and} \\
	\sigma_2 \in \E_\H (\P_\G (\sigma_1', x), k) \forall \sigma_1' \in \Sigma_\G
\end{matrix} \right.\right\}$
where $k' : Y \to S$ is defined by
\[ k' (y) = \C_\H (\sigma_2, y, k (\P_\H (\sigma_2, y))) \]
\end{itemize}

Given 4-tuples for $\G : X_1 \otimes S^* \to Y_1 \otimes R_1^*$ and $\H : X_2 \otimes S_2^* \to Y_2 \otimes R_2^*$, the tensor product
\[ \G \otimes \H : X_1 \otimes S_1^* \otimes X_2 \otimes S_2^* \to Y_1 \otimes R_1^* \otimes Y_2 \otimes R_2^* \]
is formally reduced to the type
\[ \G \otimes \H : (X_1 \times X_2) \otimes (S_1 \times S_2)^* \to (Y_1 \times Y_2) \otimes (R_1 \times R_2)^* \]
and defined by:
\begin{itemize}
	\item $\Sigma_{\G \otimes \H} = \Sigma_\G \times \Sigma_\H$
	\item $\P_{\G \otimes \H} ((\sigma_1, \sigma_2), (x_1, x_2)) = (\P_\G (\sigma_1, x_1), \P_\G (\sigma_2, x_2))$
	\item $\C_{\G \otimes \H} ((\sigma_1, \sigma_2), (x_1, x_2), (r_1, r_2)) = (\C_\G (\sigma_1, x_1, r_1), \C_\G (\sigma_2, x_2, r_2))$
	\item $\E_{\G \otimes \H} ((x_1, x_2), k) = \{ (\sigma_1, \sigma_2) \in \Sigma_1 \times \Sigma_2 \mid \sigma_1 \in \E_\G (x_1, k'_1) \text{ and } \sigma_2 \in \E_\H (x_2, k'_2) \}$
	where $k'_1 : Y_1 \to R_1$ and $k'_2 : Y_2 \to R_2$ are defined by
	\[ k'_1 (y_1) = k (y_1, \P_\H (\sigma_2, x_2)) \]
	\[ k'_2 (y_2) = k (\P_\G (\sigma_1, x_1), y_2) \]
\end{itemize}

A selection function $\varepsilon : (Y \to R) \to \mathcal P (Y)$ is lifted to a pregame $\G : X \to Y \otimes R^*$, representing a single player whose preferences are represented by $\varepsilon$.
The type is written more formally as $\G : X \otimes I^* \to Y \otimes R^*$, where $I = \{ \bullet \}$.
The definition is as follows:
\begin{itemize}
	\item $\Sigma_\G = X \to Y$
	\item $\P_\G (\sigma, x) = \sigma (x)$
	\item $\C_\G (\sigma, x, r) = \bullet$
	\item $\E_\G (x, k) = \{ \sigma : X \to Y \mid \sigma (x) \in \varepsilon (k) \}$
\end{itemize}
If $Y$ is finite and $R$ carries a rational preference relation $\preceq$, then we use the selection function
\[ \varepsilon (k) = \{ y \in Y \mid k (y) \succeq k (y') \text{ for all } y' \in Y \} \]
If instead $R = \R$ is the real numbers representing utility, we use
\[ \varepsilon (k) = \{ y \in Y \mid k (y) \geq k (y') \text{ for all } y' \in Y \} \]

A game $\G : X \otimes S^* \to Y \otimes R^*$ is called strategically trivial if $\Sigma_\G = I$, and $\E_\G (x, k) = \{ \bullet \}$ for all $x$ and $k$.
Then $\G$ is defined entirely by a function $\P : X \to Y$ and another $\C : X \times R \to S$.
Strategically trivial games are closed under both categorical composition and tensor product.
All the remaining atomic open games we need to define are strategically trivial, and so we will give only these two functions.

Let $f : X \to Y$. The covariant computation $f : X \to Y$ is defined by
\begin{itemize}
	\item $\P (x) = f (x)$
	\item $\C (x, \bullet) = \bullet$
\end{itemize}
and the contravariant computation $f^* : Y^* \to X^*$ is defined by
\begin{itemize}
	\item $\P (\bullet) = \bullet$
	\item $\C (\bullet, x) = f (x)$
\end{itemize}

The counit $\tau_X : X \otimes X^* \to I$ is defined by
\begin{itemize}
	\item $\P (x) = \bullet$
	\item $\C (x, \bullet) = x$
\end{itemize}

\end{document}